\documentclass{JHEP3}
\usepackage{multicol,multirow,graphicx}

\pdfoutput=1

\newcommand{\msun}{\,\rm M_\odot}
\newcommand{\rvir}{r_{\rm vir}}
\newcommand{\mvir}{M_{\rm halo}}
\newcommand{\Vmax}{V_{\rm max}}
\newcommand{\f}{\frac}
\newcommand{\ghalos}{GHALO$_{\rm s}$}
\newcommand{\vmin}{v_{\rm min}}

\newcommand{\be}{\begin{equation}}
\newcommand{\ee}{\end{equation}}
\newcommand{\bea}{\begin{eqnarray}}
\newcommand{\eea}{\end{eqnarray}}

\newcommand{\gsim}{ \mathop{}_{\textstyle \sim}^{\textstyle >} }

\newcommand{\kms}{{\rm ~km/s}}

\newcommand{\keV}{{\rm ~keV}}
\newcommand{\kev}{{\rm ~keV}}

\newcommand{\gev}{{\rm ~GeV}}

\title{Dark Matter Direct Detection with Non-Maxwellian Velocity Structure}

\author{Michael Kuhlen$^{a,b}$, Neal Weiner$^c$, J\"urg Diemand$^d$, Piero Madau$^e$, Ben Moore$^d$, Doug Potter$^d$, Joachim Stadel$^d$, Marcel Zemp$^f$\\
$^a$ {\rm School of Natural Science, Institute for Advanced Study, Princeton, NJ 08540}\\
$^b$ {\rm Theoretical Astrophysics Center, University of California Berkeley, Berkeley, CA 94720} \\
$^c$ {\rm Center for Cosmology and Particle Physics, Department of Physics, New York University, New York, NY 10003} \\
$^d$ {\rm Institute for Theoretical Physics, University of Zurich, 8057 Zurich, Switzerland } \\
$^e$ {\rm Department of Astronomy \& Astrophysics, University of California Santa Cruz, Santa Cruz, CA 95064} \\
$^f$ {\rm Department of Astronomy, University of Michigan, Ann Arbor, Michigan 48109} \\

E-mail: \email{mqk@astro.berkeley.edu,neal.weiner@nyu.edu}
}


\preprint{}

\abstract{The velocity distribution function of dark matter particles
  is expected to show significant departures from a Maxwell-Boltzmann
  distribution. This can have profound effects on the predicted dark
  matter - nucleon scattering rates in direct detection experiments,
  especially for dark matter models in which the scattering is
  sensitive to the high velocity tail of the distribution, such as
  inelastic dark matter (iDM) or light (few GeV) dark matter (LDM),
  and for experiments that require high energy recoil events, such as
  many directionally sensitive experiments. Here we determine the
  velocity distribution functions from two of the highest resolution
  numerical simulations of Galactic dark matter structure (Via Lactea
  II and GHALO), and study the effects for these scenarios. For
  directional detection, we find that the observed departures from
  Maxwell-Boltzmann increase the contrast of the signal and change the
  typical direction of incoming DM particles. For iDM, the expected
  signals at direct detection experiments are changed dramatically:
  the annual modulation can be enhanced by more than a factor two, and
  the relative rates of DAMA compared to CDMS can change by an order
  of magnitude, while those compared to CRESST can change by a factor
  of two. The spectrum of the signal can also change dramatically,
  with many features arising due to substructure. For LDM the spectral
  effects are smaller, but changes do arise that improve the
  compatibility with existing experiments. We find that the phase of
  the modulation can depend upon energy, which would help discriminate
  against background should it be found.}

\begin{document} 

\section{Introduction}

Direct detection experiments aim to detect the low energy nuclear
recoil from rare scattering events between dark matter (hereafter DM)
particles, assumed to be weakly interacting massive particles (WIMPs),
and target nuclei. The event rate and its energy spectrum depend on
the properties of the DM distribution at Earth's location, about 8.5
kpc from the Galactic Center \cite{Kerr:1986hz}. Typical calculations
of the scattering rate assume a ``standard halo model'' (SHM)
consisting of a local DM density of $0.3 \pm 0.1$ GeV cm$^{-3}$ and a
Maxwell-Boltzmann (MB) velocity distribution function with a
(three-dimensional) dispersion of $270 \pm 70$ km/s
\cite{Kamionkowski1998} truncated at an escape speed of $\sim 550$
km/s. Recent numerical simulations of the formation of Galactic-scale
DM halos have reached the necessary resolution to directly test these
assumptions.

At the same time, absent a positive signal, a set of uniform halo
assumptions allows a simple means to compare different
experiments. However, in light of the recent results from DAMA/LIBRA
\cite{Bernabei2008}, confirming earlier results from DAMA/NaI
\cite{Bernabei:2000qi}, it is important to consider halo model
uncertainties when discussing exclusion limits from experiments with
different targets, or energy ranges. This is particularly important
because proposals such as light dark matter (LDM)
\cite{Bottino:2003cz,Gondolo:2005hh} and inelastic dark matter (iDM)
\cite{Smith:2001hy,Tucker-Smith:2004jv}, which aim to reconcile DAMA
with null results from other experiments, sample the high velocity
component of the WIMPs preferentially, and it is especially here that
the Maxwell-Boltzmann distribution is expected to break down.

In the hierarchical structure formation paradigm of standard cold dark
matter cosmology the DM halo of a typical galaxy is built up through
the merger of many individual gravitationally bound progenitor halos,
which themselves were assembled in a hierarchical fashion. High
resolution cosmological simulations, such as Via Lactea
\cite{Diemand2007,Diemand2008}, GHALO \cite{Stadel2009}, and Aquarius
\cite{Springel2008}, have shown that this merging process is in
general incomplete, with the dense cores of many of the merging halos
surviving as subhalos orbiting within their respective host halos.
During pericenter passages tidal forces can strip off a large fraction
of a subhalo's material, but the resulting cold tidal streams can
readily be identified as velocity space substructure.  The resulting
DM halos are not perfectly phase mixed, and the assumption of a smooth
halo is in general not a good one, neither in configuration space nor
in velocity space \cite{Zemp2009}.

At $8.5$ kpc the Sun is located quite close to the Galactic Center, at
least when compared to the overall extent of the Milky Way's DM halo
($\rvir \sim 200 - 300$ kpc). This central region is notoriously
difficult for cosmological numerical simulations to resolve, as very
high particle numbers and very short time steps are required to avoid
the so-called ``over-merging problem'' \cite{Klypin:1997fb}, which has
until recently resulted in an artifically smooth central halo devoid
of any substructure. With the advent of $\mathcal{O}(10^9)$ particle
simulations at the Galactic scale this problem finally seems to have
been overcome, with hundreds of subhalos identified at $\lesssim 20$
kpc. Nevertheless the local phase space structure is far from
completely resolved, and likely never will be through direct numerical
simulation. Any estimation of the importance of local density or
velocity substructure based on cosmological simulation must thus rely
on extrapolations over many orders of magnitude below its resolution
limit.

Local density variations due to the clumpiness of the DM halo are
unlikely to significantly affect the direct detection scattering
rate. Based on the Aquarius Project suite of numerical simulations,
Vogelsberger et al. (2008) \cite{Vogelsberger2009} report that at more
than 99.9\% confidence the DM density at the Sun's location
differs by less than 15\% from the average over a constant density
ellipsoidal shell. Extrapolating from their numerical convergence
study they estimate a probability of $10^{-4}$ of the Sun residing in
a bound subhalo of any mass.  Analytical work by Kamionkowski \&
Koushiappas (2008) \cite{Kamionkowski2008} predicts a positively
skewed density distribution with local densities as low as one tenth
the mean value, but probably not much less than half.

The situation for velocity substructure is less clear.  It is well
established that numerically simulated dark matter halos exhibit
significant velocity anisotropy and global departures from a
Maxwell-Boltzmann distribution \cite{Diemand2004, Wojtak2005,
  Hansen:2005yj, Hansen2006}. The implications for direct detection
experiments of these global departures from the standard Maxwellian
model have previously been investigated in the context of a standard
WIMP model \cite{Moore2001, Savage:2006qr, Vergados2008, Fairbairn2009} and for
inelastic dark matter \cite{March-Russell2009, Ling2009}, and they
were found to result in appreciable differences (factor of a few) in
the total event rates and the annual modulation signal. Most recently
Vogelsberger et al. (2008) reported significant structure
(``wiggles'') in the velocity distribution function measured in their
high-resolution Aquarius simulations, which they attributed to events
in the halo's mass assembly history. Their analysis concluded that
velocity substructure due to bound subhalos or unbound tidal streams,
however, does not influence the detector signals, since it makes up a
highly sub-dominant mass fraction locally.

The aim of this paper is take a closer look at this velocity space
substructure and to examine its impact on the direct detection signal
for models that are particularly sensitive to the high velocity tail,
such as LDM or iDM. In contrast to Vogelsberger et al. (2008), we find
that both global and local departures from the best-fit
Maxwell-Boltzmann distribution can significantly affect the total
event rate, the annual modulation, and the recoil energy
spectrum. Parameter exclusion limits derived using a standard MB halo
model are likely to be overly restrictive.

This paper is organized as follows: in Section~\ref{sec:sims} we
present velocity distribution functions derived from the
high-resolution numerical simulations Via Lactea II and GHALO. In
Section~\ref{sec:directional} we look at the implications of high
velocity substructure for direct detection experiments with
directional sensitivity. In Section~\ref{sec:dm_models} we consider
iDM and LDM models and show how the observed local and global
departures from the MB model affect scattering event rates and recoil
spectra at several ongoing direct detection experiments, and how this
modifies parameter exclusion limits. A summary and discussion of our
results can be found in Section~\ref{sec:discussion}.

\section{Results from Numerical Simulations}
\label{sec:sims}

The nuclear scattering event rate depends on the size of the detector,
the type of target material, the scattering cross section, and the
number density and velocity distribution of the impinging DM
particles. We defer calculations of the expected event rate for
various experimental setups and types of DM models to
section~\ref{sec:dm_models}, and focus in this section on the particle
velocity distributions, which we determine directly from numerical
simulations.

\subsection{The Via Lactea and GHALO simulations}

Our analysis is based on two of the currently highest resolution
numerical simulations of Galactic DM structure: Via Lactea II (VL2)
\cite{Diemand2008} and GHALO \cite{Stadel2009}. Both are cosmological
cold DM N-body simulations that follow the hierarchical growth and
evolution of a Milky-Way-scale halo and its substructure from initial
conditions in the linear regime ($z=104$ for VL2, $z=58$ for GHALO)
down to the present epoch. For details about the setup of the
simulations we refer the reader to the above references. The VL2 host
halo is resolved with $\sim 400$ million particles of mass $m_p=4,100
\msun$ within its virial radius\footnote{$\rvir$ is defined as the
  radius enclosing a density of $\Delta_{\rm vir}=389$ times the
  background density.} of $\rvir=309$ kpc and has a mass of $\mvir=1.7
\times 10^{12} \msun$ and peak circular velocity $\Vmax=201.3$
km/s. The GHALO host is somewhat less massive, $\mvir=1.1 \times
10^{12} \msun$ and $\Vmax=152.7$ km/s, but even more highly resolved,
with 1.1 billion particles of mass $m_p=1,000 \msun$ within its
$\rvir=267$ kpc. For reference we show the circular velocity of the
two halos in Fig.\ref{fig:vc_profiles}. In order to facilitate a more
direct comparison between the two halos, we have also scaled GHALO to
match VL2's $\Vmax$ by multiplying the simulation's length and
velocity units by a factor $f=\Vmax({\rm VL2})/\Vmax({\rm
  GHALO})=1.32$, and the mass unit by $f^3$. We refer to this model as
\ghalos. The circular velocity of these three halos at 8.5 kpc is
158.1, 121.7, and 148.9 km/s for VL2, GHALO, and \ghalos,
respectively.

\FIGURE[t]{
\includegraphics[width=0.6\textwidth]{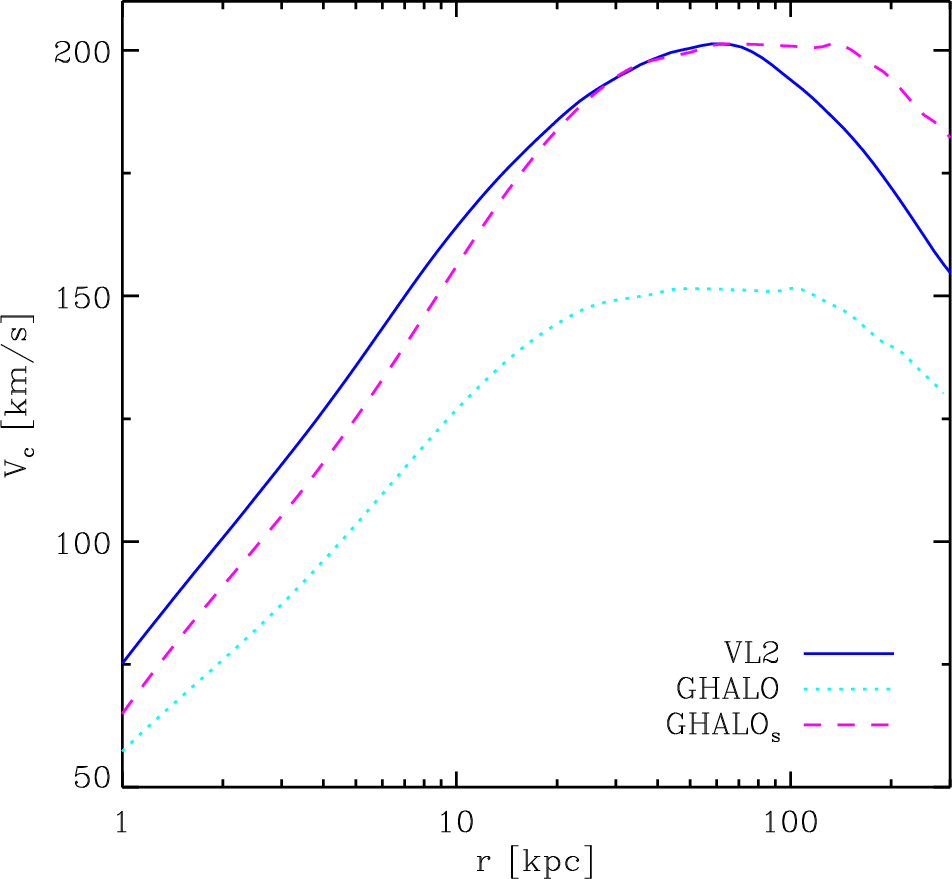}
\caption{Circular velocity profiles of the VL2, GHALO, and \ghalos\ host halos.}
\label{fig:vc_profiles}}

\subsection{Velocity Modulus Distributions}
\label{sec:vdist}

The DM-nucleon scattering event rate is directly proportional to
\begin{equation}
g(\vmin) = \int_{\vmin}^\infty \f{f(v)}{v}dv,
\label{eq:gvmin}
\end{equation}
where $f(v)$ is the DM velocity distribution function in the Earth's
rest frame and $\vmin(E_R)$ is the minimum velocity that can result in
a scattering with a given nuclear recoil energy $E_R$. For a target
with nuclear mass $m_N$ and a WIMP/nucleon reduced mass $\mu=m_N
m_\chi/(m_N + m_\chi)$, $\vmin(E_R)$ is given by
\begin{equation}
\left(\f{\vmin}{c}\right)^2 = \f{1}{2} \f{m_N E_R}{\mu^2} \left(1 + \f{\mu}{m_N E_R} \delta \right)^2.
\label{eq:vmin_Erecoil}
\end{equation}
The $\delta$ refers to the possible mass splitting between the
incoming and outgoing DM particle, which would be 0 for standard and
light DM and $\mathcal{O}$(100 keV) for inelastic DM.

We determine $f(v)$ in the halo rest frame directly from the particle
velocities in our numerical simulations, and in the Earth's rest frame
by first applying a Galilean velocity boost by $v_\oplus(t)$. The
Earth's velocity with respect to the Galactic center is the sum of the
local standard of rest (LSR) circular velocity around the Galactic
center, the Sun's peculiar motion with respect to the LSR, and the
Earth's orbital velocity with respect to the Sun,
\begin{equation}
\vec{v}_\oplus(t) = \vec{v}_{\rm LSR} + \vec{v}_{\rm pec} + \vec{v}_{\rm orbit}(t).
\end{equation}
We follow the prescription given in Chang et al. (2008)
\cite{Chang:2008gd} and set $\vec{v}_{\rm LSR} = (0, 220, 0)$
km/s, $\vec{v}_{\rm pec} = (10.00, 5.23, 7.17)$ km/s
\cite{Dehnen1998}, and $\vec{v}_{\rm orbit}(t)$ as specified in
reference \cite{Lewin1996}. The velocities are given in the
conventional $(U,V,W)$ coordinate system where $U$ refers to motion
radially inwards towards the Galactic center, $V$ in the direction of
Galactic rotation, and $W$ vertically upwards out of the plane of the
disk. We associate these three velocity coordinates with the $(v_r,
v_\theta, v_\phi)$ coordinates of the simulation particles. 

The Earth's orbital motion around the Sun results in the well-known
annual modulation of the scattering rate, which the DAMA collaboration
claims to have detected at very high statistical significance
\cite{Bernabei2008}. For the SHM the peak of this modulation occurs
around June 2$^{\rm nd}$, when the Earth's relative motion with
respect to the Galactic DM halo is maximized.

\FIGURE[t]{
\includegraphics[width=\textwidth]{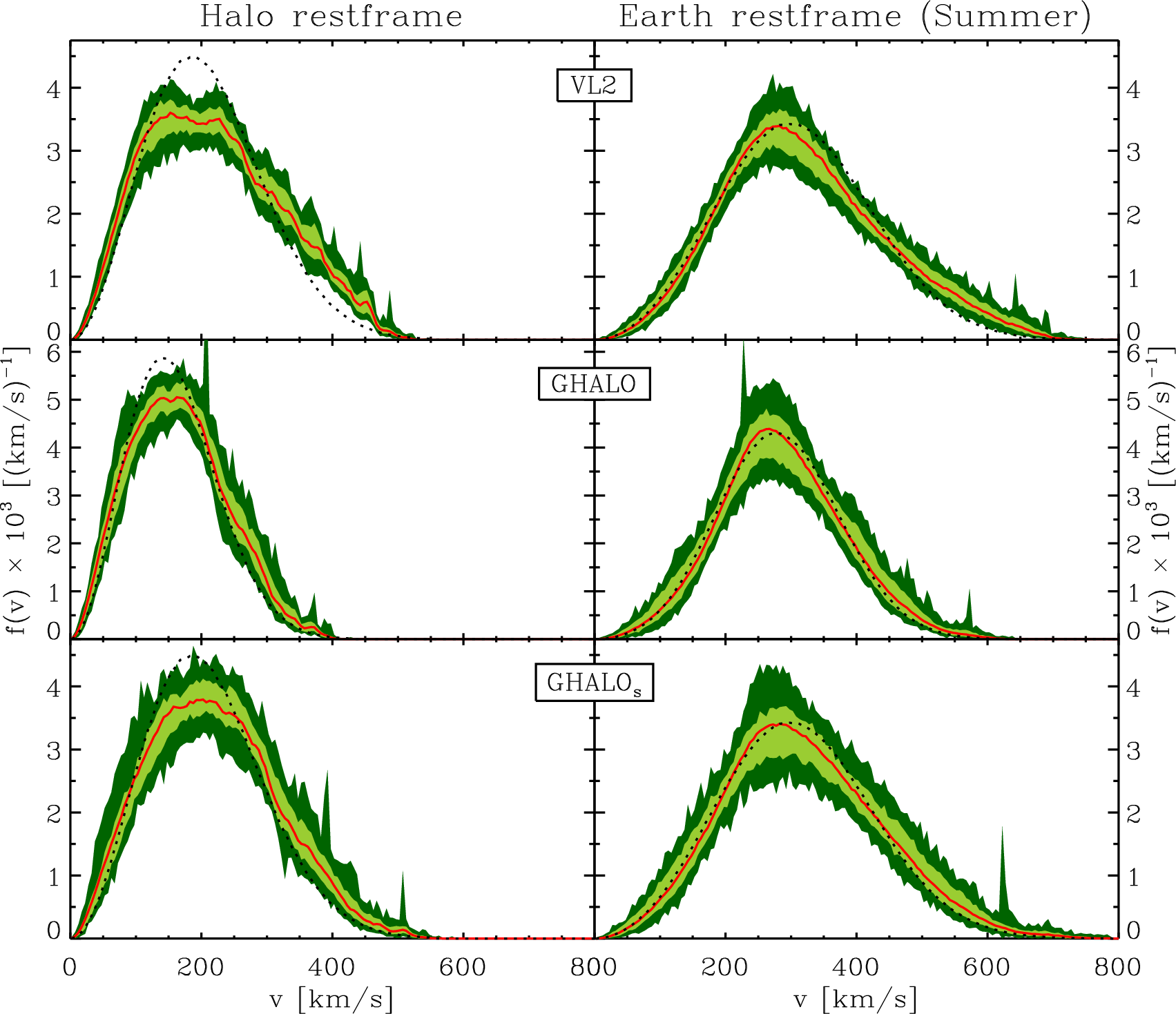}
  \caption{Velocity distribution functions: the left panels are in the
    host halo's restframe, the right panels in the restframe of the
    Earth on June 2$^{\rm nd}$, the peak of the Earth's velocity
    relative to Galactic DM halo. The solid red line is the
    distribution for all particles in a 1 kpc wide shell centered at
    8.5 kpc, the light and dark green shaded regions denote the 68\%
    scatter around the median and the minimum and maximum values over
    the 100 sample spheres, and the dotted line represents the
    best-fitting Maxwell-Boltzmann
    distribution.}\label{fig:fv_distributions}}

We have measured the DM velocity distribution from all particles in a
1 kpc wide spherical shell (8 kpc $< r <$ 9 kpc), containing 2.1, 5.4,
and 3.6 million particles in VL2, GHALO, and \ghalos, respectively.
The large particle numbers in these measurements result in a very
small statistical uncertainty, but fail to capture any local
variations. To address this we have also determined $f(v)$ from the
particles in 100 randomly distributed sample spheres centered at 8.5
kpc. These sample spheres have radii of 1.5 kpc for VL2 and 1 kpc for
GHALO and \ghalos, and contain a median of 31,281, 21,740, and 14,437
particles in the three simulations.\footnote{Tables of $g(\vmin)$
  determined from the spherical shell and the 100 sample spheres, and
  tracing the annual modulation over 12 evenly spaced output times,
  are available for download at
  \texttt{http://astro.berkeley.edu/$\sim$mqk/dmdd/}.}

The resulting distributions, both in the halo rest frame and
translated into Earth's rest frame, are shown in
Fig.~\ref{fig:fv_distributions}. The shell averaged distribution is
plotted with a solid line, while the light and dark green shaded
regions indicate the 68\% scatter around the median and the absolute
minimum and maximum values of the distribution over the 100 sample
spheres. For comparison we have also overplotted the best-fitting
Maxwell-Boltzmann (hereafter MB) distributions, with 1D velocity
dispersion of $\sigma_{\rm 1D} =$ 130, 100, and 130
km/s. These clearly underpredict both the low and high velocity tails
of the actual distribution. This is not a new result and has
previously been found in cosmological numerical simulations
\cite{Diemand2004, Wojtak2005, Hansen:2005yj,
  Vogelsberger2009}. Actually there is no reason to assume that a
self-gravitating, dissipationless system would have a locally
Maxwellian velocity distribution, and in fact it has been shown that
self-consistent, stable models of cuspy DM structures require just
such non-Gaussianity \cite{Hernquist1993, Kazantzidis2004}.

In addition to its overall non-Maxwellian nature, we notice several
broad bumps present in both the shell averaged and, at very similar
speeds, in the sub-sample $f(v)$. Similar features were reported
by Vogelsberger et al. (2008) \cite{Vogelsberger2009} for the host
halos of their completely independent Aquarius simulations, and thus
appear to be robust predictions of hierarchically formed collisionless
objects. Vogelsberger et al. also showed that the broad bumps are
independent of location and persistent in time and hence reflect the
detailed assembly history of the host halo, rather than individual
streams or subhalos. The extrema of the sub-sample distributions,
however, exhibit numerous distinctive narrow spikes at certain
velocities, and these are due to just such discrete structures. Note
that although only a small fraction of sample spheres exhibits such
spikes, they are clearly present in some spheres in all three
simulations. The Galilean transform into the Earth's rest frame washes
out most of the broad bumps, but the spikes remain visible, especially
in the high velocity tails, where they can profoundly affect the
scattering rates for inelastic and light DM models (see
Section~\ref{sec:dm_models}).

In order to assess the dependence of these features on the sample
sphere size, we also considered for the VL2 simulation sphere radii of
1 and 2 kpc, containing a median of 9,200 and 74,398 particles,
respectively. The coarse features of the distributions persist, but of
course the prevalence of the spikes increases with sample sphere size,
as more of the substructure is probed. It is difficult to assess with
our simulations the true likelihood of significant local velocity
substructure, as it depends on the abundance and physical extent of
subhalos and tidal streams many orders of magnitude below the length
scales that we can accurately resolve. Higher resolution numerical
simulations, as well as analytical models \cite{Afshordi2009},
perhaps in conjunction with simulations \cite{Vogelsberger2008}, will
be necessary to settle this question.

\subsection{The effects of neglected baryonic physics}
The values of $\sigma_{\rm 1D}$ we report here may appear surprisingly
low to a reader familiar with the standard isothermal MB halo
assumption of $\langle v^2 \rangle = 3 \, \sigma_{\rm 1D}^2 = 3/2 \, v_0^2$,
where $v_0$, the peak of the MB distribution, i.e. the most probable
speed, is assumed to be equal to the rotation velocity of the Sun
around the galaxy, $v_0 \simeq 220$ km/s. In this standard model
$\sigma_{\rm 1D}$ would be 156 km/s, considerably higher than our
values of 130 km/s and 100 km/s, respectively. In fact, the local
circular velocity $v_c$ and velocity dispersion $\sigma$ are only
indirectly related and not necessarily equal. While $v_c$ is set by
the local radial gradient of the potential, $\sigma$ depends on the
shape of the potential at exterior radii.  For a non-rotating
spherical system the relation between $v_c(r)$ and the radial velocity
dispersion $\sigma_r(r)$ is given by
\begin{equation}
v_c^2 = - \sigma_r^2 \left( \f{d\ln \rho}{d\ln r} + \f{d\ln \sigma_r^2}{d\ln r} + 2\beta \right),
\end{equation}
where $\f{d\ln \rho}{d\ln r} \equiv \gamma(r)$ is the logarithmic
slope of the density profile and $\beta(r) \equiv 1 -
\sigma_\theta^2/\sigma_r^2$ is the velocity anisotropy.  In a singular
isothermal sphere ($\gamma=-2$, $\f{d\ln \sigma_r^2}{d\ln r} = 0$,
$\beta=0$) we have $v_c = v_0$, but for an NFW profile $v_c/v_0 \approx
0.88$ at $r=r_s/2$. In the VL2 and GHALO host halos the
relation is $v_c/v_0 = 0.85$ and $0.86$, respectively.

A central baryonic condensation in the form of a Galactic disk and
bulge will deepen the central potential, raise the local circular
velocity to $\sim 220$ km/s, and increase the velocity dispersion of
DM particles at the Sun's location. Since the VL2 and GHALO
simulations do not include baryonic physics, it is not surprising that
the values of $v_c$ and $v_0$ at 8.5 kpc are lower than appropriate
for our Milky Way galaxy. The main focus of our work here is to
investigate the effects of global and local variations from the MB
assumption, and therefore we compare our results to the best-fitting
MB model ($v_0 = 184$ km/s for VL2 and \ghalos\ and 141 km/s for
GHALO) instead of the standard MB halo model ($v_0=220$ km/s).  In
principle we could have scaled just the velocities up to give
$v_0=220$ km/s, but this would remove many of the effects of
substructure by pushing it above the escape velocity. Thus, we use VL2
and \ghalos\ simulations for the parameter exclusions plots in
Section~\ref{sec:dm_models}, but it should be recognized, that the
velocity dispersion is somewhat lower than in conventional MB halo
parameterizations.

\subsection{Local Escape Speed}

The escape speed from the Sun's location in the Galactic halo is
another factor that can strongly affect scattering rates, especially
for inelastic and light DM models. The RAVE survey's sample of
high-velocity stars constrains the Galactic escape velocity to lie
between 498 and 608 km/s at 90\% confidence, with a median likelihood
of 544 km/s \cite{Smith2007}. This is in good agreement with the
highest halo rest frame speed of any particle in our 8.5 kpc spherical
shells, namely 550 km/s in VL2 and 586 km/s in \ghalos. The lower
$\Vmax$ of the GHALO host is reflected in a significantly lower escape
speed, only 433 km/s. The corresponding maximum speeds in the Earth
rest frame are 735-761, 773-802, and 634-660 km/s, where the range
refers to the modulation introduced by the Earth's orbit around the
Sun.

\subsection{Radial and Tangential Distributions}

\FIGURE[t]{
\includegraphics[width=0.7\textwidth]{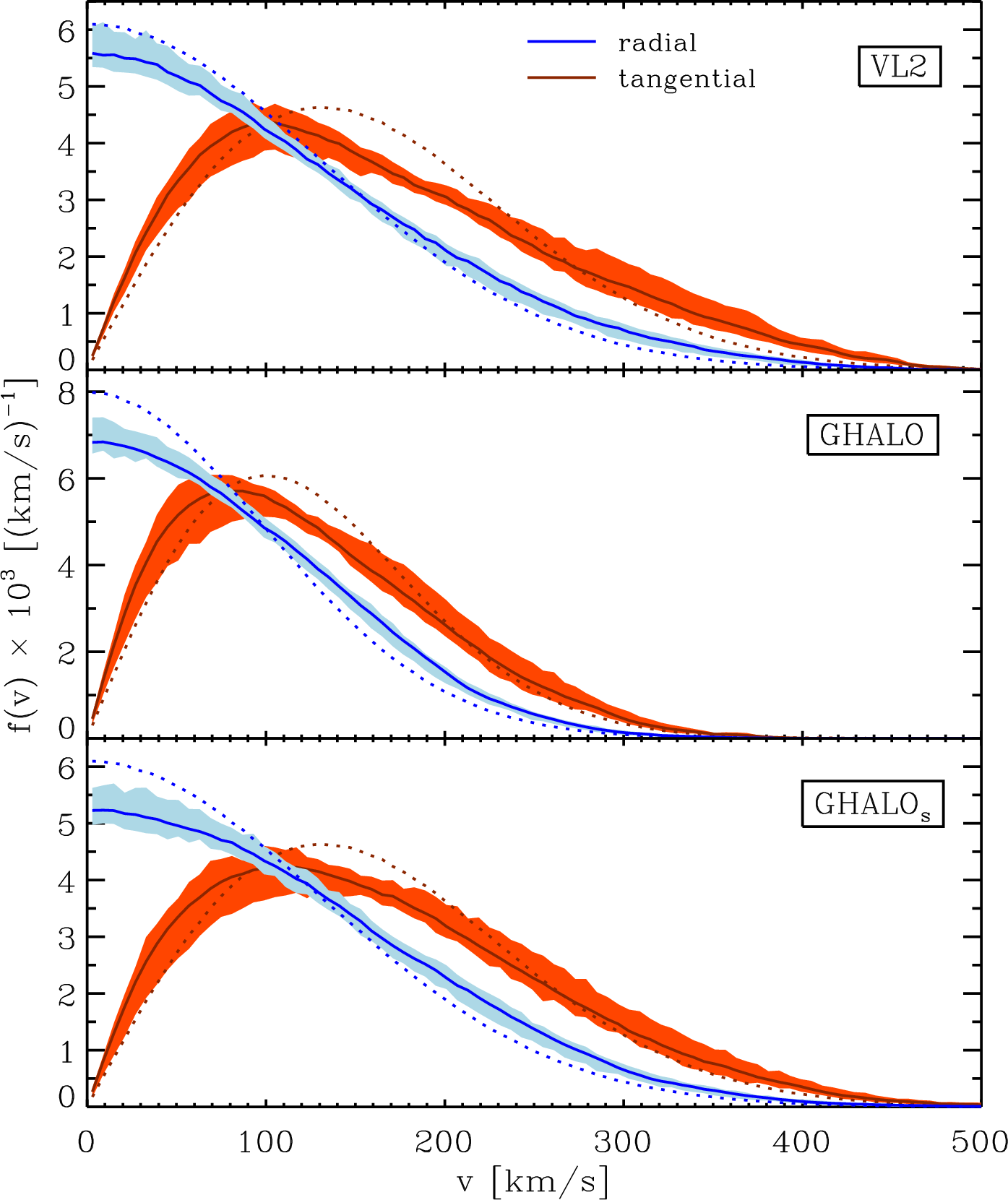}
\caption{Radial and tangential velocity distribution functions. The
  solid lines show the shell average, the shaded range the 68\%
  scatter around the median, and the dotted curve shows the best-fit
  MB model.}
\label{fig:vr_vtan}}

For comparison with past work \cite{March-Russell2009,Fairbairn2009}
we also present separately the distribution functions of the radial
$v_r$ and tangential $v_t = \sqrt{v_\theta^2 + v_\phi^2}$ velocity
components in Fig.~\ref{fig:vr_vtan}. We fit these distributions with
functions like the ones in Fairbairn \& Schwetz (2009)
\cite{Fairbairn2009}, except that we don't normalize the velocities by
the square root of the gravitational potential at the particles'
location and instead provide separate fits for each simulation. We
also include an estimate of the variance due to local velocity
substructure by separately fitting the distribution in each of the
sample spheres. The fitting functions are
\begin{eqnarray}
f(v_r) & = & \f{1}{N_r} \exp\left[ - \left( \f{v_r^2}{\bar{v}_r^2} \right)^{\alpha_r} \right] \label{eq:vr_fit} \\
f(v_t) & = & \f{v_t}{N_t} \exp\left[ - \left( \f{v_t^2}{\bar{v}_t^2} \right)^{\alpha_t} \right] \label{eq:vt_fit},
\end{eqnarray}
and the parameters of these fits are listed in Table~\ref{tab:fits}.
The normalizations $N_r$ and $N_t$ can readily be obtained numerically
for a given set of parameters by ensuring that the distributions
integrate to unity.

\TABULAR{cc|cccc|cccc}{
 & & \multicolumn{4}{c}{{\Large radial}} & \multicolumn{4}{|c}{{\Large tangential}} \\
 & & shell & median & $16^{\rm th}$ & $84^{\rm th}$ & shell & median & $16^{\rm th}$ & $84^{\rm th}$ \\
\hline
\multirow{2}{*}{VL2} & $\bar{v}_{r,t}$ [km/s] & 202.4 &  199.9 & 185.5 & 212.7 & 128.9 &  135.1 & 124.2 & 148.9 \ \\
                     & $\alpha_{r,t}$ & 0.934 &  0.941 & 0.877 & 0.985 & 0.642 &  0.657 & 0.638 & 0.674 \\
\hline
\multirow{2}{*}{GHALO} & $\bar{v}_{r,t}$ [km/s] & 167.9 &  163.6 & 156.4 & 173.0 & 103.1 &  114.3 &  93.21 & 137.0 \ \\
                     & $\alpha_{r,t}$ & 1.12 &  1.11 & 1.02 & 1.20 & 0.685 &  0.719 & 0.666 & 0.819 \\
\hline
\multirow{2}{*}{\ghalos} & $\bar{v}_{r,t}$ [km/s] & 217.9 &  213.8 & 202.3 & 226.6 & 138.2 &  162.2 & 125.1 & 183.1 \ \\
                     & $\alpha_{r,t}$ & 1.11 &  1.11 & 1.01 & 1.18 & 0.687 &  0.759 & 0.664 & 0.842 \\
}{Radial and tangential velocity distribution fit parameters (see Eqs. \protect \ref{eq:vr_fit} and \protect \ref{eq:vt_fit}). The columns labeled ``shell'' refer to the fit for all particles in the 1 kpc wide shell centered on 8.5 kpc, whereas the following three columns give the median as well as the $16^{\rm th}$ and $84^{\rm th}$ percentile of the distribution of fits over the 100 sample spheres (see text for more detail).}\label{tab:fits}

\subsection{Modulation Amplitude and Peak Day}

The annual modulation of the scattering rate $g(\vmin)$
(Eq.~\ref{eq:gvmin}) grows with $\vmin$, since a reduction in the
number of particles able to scatter makes the summer-to-winter
difference relatively more important. At sufficiently high velocities
the modulation amplitude can even reach unity, when during the winter
there simply aren't any particles with velocities above $\vmin$.

The expected modulation amplitude for a given experiment is then
determined by the relation between the measured recoil energy $E_R$
and $\vmin$, see Eq.~\ref{eq:vmin_Erecoil}. There are important
qualitative difference between standard DM models ($\delta=0$) and
inelastic models ($\delta \sim 100$ keV). Inelastic DM models require
a higher $\vmin$ for a given $E_R$, resulting in a lower total
scattering rate and a more pronounced modulation.  Furthermore, while
for standard DM $\vmin$ grows with $E_R$, in inelastic models $\vmin$
typically falls with $E_R$ (for $\delta > E_R m_N/\mu$).

\FIGURE[t]{
\includegraphics[width=\textwidth]{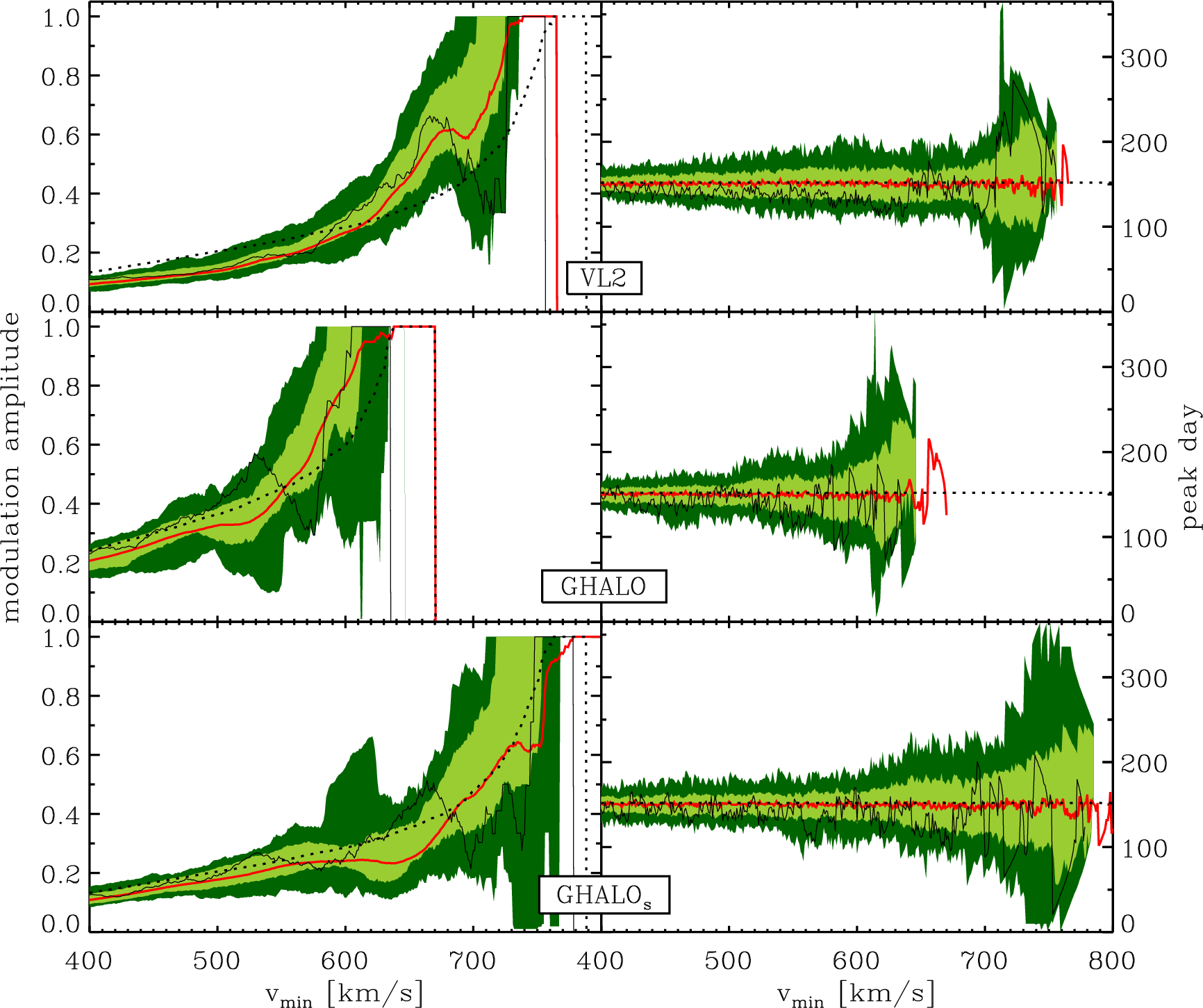}
  \caption{The amplitude (left panels) and peak day (right) as a
    function of $\vmin$. The solid red line indicates the shell
    averaged quantities, and the dotted line the best-fit MB
    distribution. The light and green shaded regions cover the central
    68\% region around the median and the minimum and maximum values
    of the distribution over the 100 sample spheres. The thin black
    line shows the behaviour of one example sample sphere.}
\label{fig:amplitude}}

In the left panels of Fig.~\ref{fig:amplitude} we show the $\vmin$
dependence of the annual modulation amplitude, $({\rm
  max}(g(\vmin)) - {\rm min})/({\rm max} + {\rm min})$
as measured in our simulations. We focus on
the high $\vmin$ region that is relevant for inelastic and light dark
matter models, as well as directional detection experiments. In VL2 and \ghalos\ the shell averaged modulation
amplitude (solid red line) rises from about 20\% at $\vmin=400$ km/s
to unity at $\sim 750$ km/s. The GHALO amplitudes are shifted to lower
$\vmin$, growing from 40\% at 400 km/s to 100\% already at $\sim 600$
km/s. The strong high velocity tails of $f(v)$
(Fig.~\ref{fig:fv_distributions}) result in somewhat lower modulation
amplitudes compared to the best-fit MB distributions (dotted line).
At the very highest velocities $f(v)$ drops below the Maxwellian
distribution in VL2 and GHALO, and this leads to the rise in
amplitudes above the Maxwellian case for $\vmin > 600$ and 550 km/s,
respectively. In \ghalos\ the distribution more closely follows the
MB fit, and only barely rises above it at $\vmin > 670$ km/s.
Interestingly, all three simulations exhibit a pronounced dip in the
modulation amplitude at close to the highest $\vmin$. These correspond
to bumps in $f(v)$ discussed in Section~\ref{sec:vdist}.

As before, the light and dark shaded green regions in
Fig.~\ref{fig:amplitude} cover the 68\% region around the median and
the extrema of the distribution over the 100 sample spheres.  Over
most of the range of $\vmin$, the typical modulation amplitude in a
sample sphere differs from the spherical shell average by less than
10\%. The extrema of the distribution, however, can differ much more
significantly, especially at higher velocities. If the Sun happens to
be passing through a fast moving subhalo or tidal stream, the typical
velocity of impinging DM particles can greatly differ from the smooth
halo expectation, leading to an increase (or decrease) in the overall
scattering rate and modulation amplitude. For a conventional massive
($>10$ GeV) DM particle, scattering events are dominated by the peak
of the velocity distribution function and the effects from streams or
subhalos are washed out and typically negligible
\cite{Vogelsberger2009}. However, whenever scattering events are
dominated by particles on the high velocity tail of the distribution,
either because of a velocity threshold (inelastic DM), a particularly
low particle mass (light DM), or for detectors intrinsically sensitive
to only high recoil energies (e.g. many directionally sensitive
detectors), the presence of velocity substructure can have a profound
impact on the event rates.

Another potentially interesting signature of velocity space structure
is the shift in the peak day of the annual modulation, as was explored in \cite{Savage:2006qr}. For an
isotropic velocity distribution (e.g. MB) the peak day is independent
of $\vmin$ and occurs around the beginning of June. Any kind of
departure from isotropy, however, would leave its signature as a
change in the phase of the modulation. In the most extreme case of a
very massive DM stream moving in exactly the same direction as the
Sun, the phase of modulation could flip completely, with the maximum
scattering amplitude occurring in the winter. 

In the right panel of Fig.~\ref{fig:amplitude} we show the $\vmin$
dependence of the peak day as measured in our simulations. The
spherical shell average is consistent with a phase shift of zero,
except at the very highest velocities where a few discrete velocity
structures dominate the shell average and introduce small velocity
dependent phase shifts. The peak days in the sample spheres, however,
often differ by $\sim 20$ days from the fiducial value.  As an example
we have plotted a curve for one of the sample spheres. The peak day
determination appears to be quite noisy here, but since the particle
numbers are still fairly large ($N(>\vmin) = 9114, 802, 105$ for
$\vmin=400, 600, 670$ km/s in the VL2 sample sphere shown), we believe
that the variations in peak day arise from actual velocity structure
rather than discreteness noise. The 68\% scatter of the phase shifts
over the 100 sample spheres remains remarkable constant at $\pm 10-20$
days throught most of the $\vmin$ range. At higher velocities ($\vmin
> 600 - 700$ km/s) the typical phase shifts grow to $\sim 50$ days,
which could be due to the increasing relative importance of individual
streams, but may also be due to particle discreteness noise.
\textit{Based on our measurements we would expect some amount of
  variation in the peak day as a function recoil energy. As it is hard
  to imagine any background contamination to exhibit such a phase
  shift, this raises the possibility of such a measurement confirming
  a DM origin of the DAMA modulation signal.}

We conclude this section by noting that our simulations reveal both
global (shell averaged) and local (sample spheres) departures from the
standard halo model in velocity space. These can have a significant
effects on the overall scattering rate, as well as on the amplitude
and peak day of the annual modulation thereof. In the next section we
go on to explore how these effects translate into predicted detection
rates for actual direct detection experiments, and how they modify
parameter exclusions plots based on existing null-detections.

\section{Velocity Substructure and Directional Detection}
\label{sec:directional}

Up to this point we have focused on the integrated features of the
halo, i.e., $f(v)$ or $g(\vmin)$, but there are many more structures
that appear that are not pronounced in these measures. In particular,
the presence of clumps and streams, while contributing to the bumps
and wiggles of these functions, can be more pronounced when the
direction of the particles' motions, rather than just the amplitude,
is considered. These features are especially important for directional
WIMP detectors, such as DRIFT \cite{SnowdenIfft2003}, NEWAGE
\cite{Nishimura2009} and DMTPC \cite{Sciolla2009}, where the presence
of such structures could show up as a dramatic signal. Although we do not discuss it here, such directional experiments have been argued to be especially important for testing inelastic models \cite{Finkbeiner:2009ug,Lisanti:2009vy}.

\FIGURE[t]{
  \includegraphics[width=\textwidth]{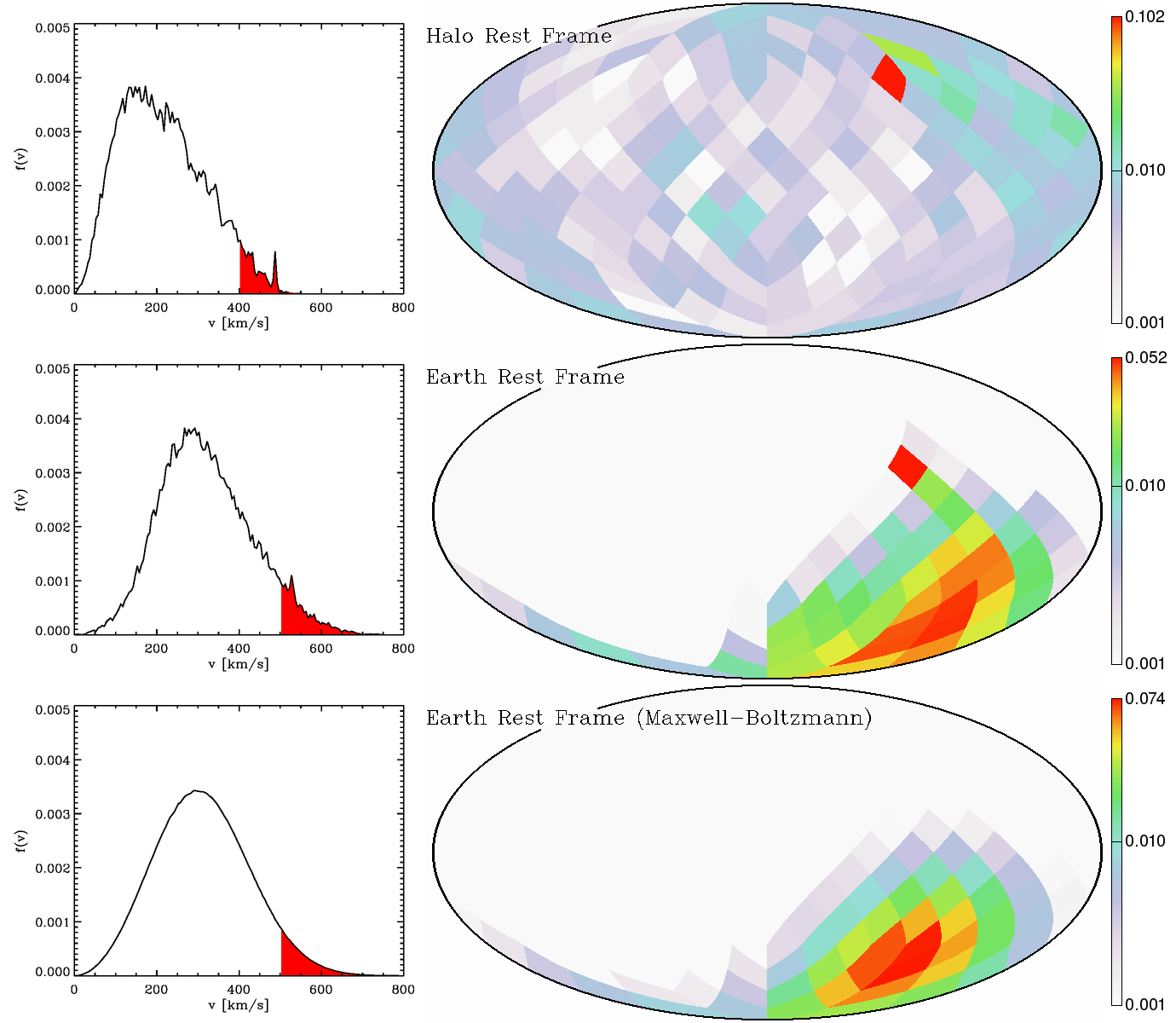}
  \caption{$f(v)$ and HealPix skymaps of the fraction of particles
    above $\vmin$ coming from a given direction, for VL2 sample sphere
    \#03 which contains a fast-moving subhalo. The top row is in the
    halo rest frame ($\vmin=400$ km/s), and the middle translated into
    the Earth rest frame ($\vmin=500$ km/s). For comparison the bottom
    row shows the Maxwell-Boltzmann halo case without substructure.}
\label{fig:sphere03}}

To quantify this, we begin by searching for ``hotspots'' on the
sky. To do this, we make a map of the sky in HealPix, and consider the
flux of WIMPs from each direction in the sky. We divide the sky into
192 equal regions of 215 square degrees, and determine $p_i$, the
fraction of particles above $\vmin$ with a velocity vector pointing
towards bin $i$. We take the same 100 sample spheres as before, but
note that beyond determining sample sphere membership we do not
consider the location of a particle: the assignment to a given sky
pixel is based solely on the direction of its velocity vector. All of
the structure in a given sample sphere is considered \textit{local},
i.e. able to influence the signal at an Earth-bound direct detection
experiment. 

As an example of the kind of effects high velocity substructure can
produce, we show in Fig.~\ref{fig:sphere03} the speed distribution
$f(v)$ and skymaps for VL2 sample sphere \#03. In the halo rest frame
(top row, $\vmin=400$ km/s) a very pronounced feature is visible due
to the presence of a subhalo moving with galacto-centric velocity
modulus of $\sim 440$ km/s. This feature persists in the Earth rest
frame (center row, $\vmin=500$ km/s), where the direction of the
subhalo's motion is ``hotter'' than the ``DM headwind'' hotspot in the
direction of Earth's motion. When translating to the Earth's rest
frame, there is an additional degree of freedom arising from the
unspecified plane of the Galactic disk. We associate radial motion
towards (away from) the Galactic Center with the center (anti-center)
of the map, but the hotspot arising from Earth's motion is free to be
rotated around this axis. In this example we have chosen this rotation
to maximize the angle between the subhalo hotspot and Earth's
motion. In the bottom row we show for comparison the Earth rest frame
map for the MB-case without any substructure.

\FIGURE[t]{
  \includegraphics[width=\textwidth]{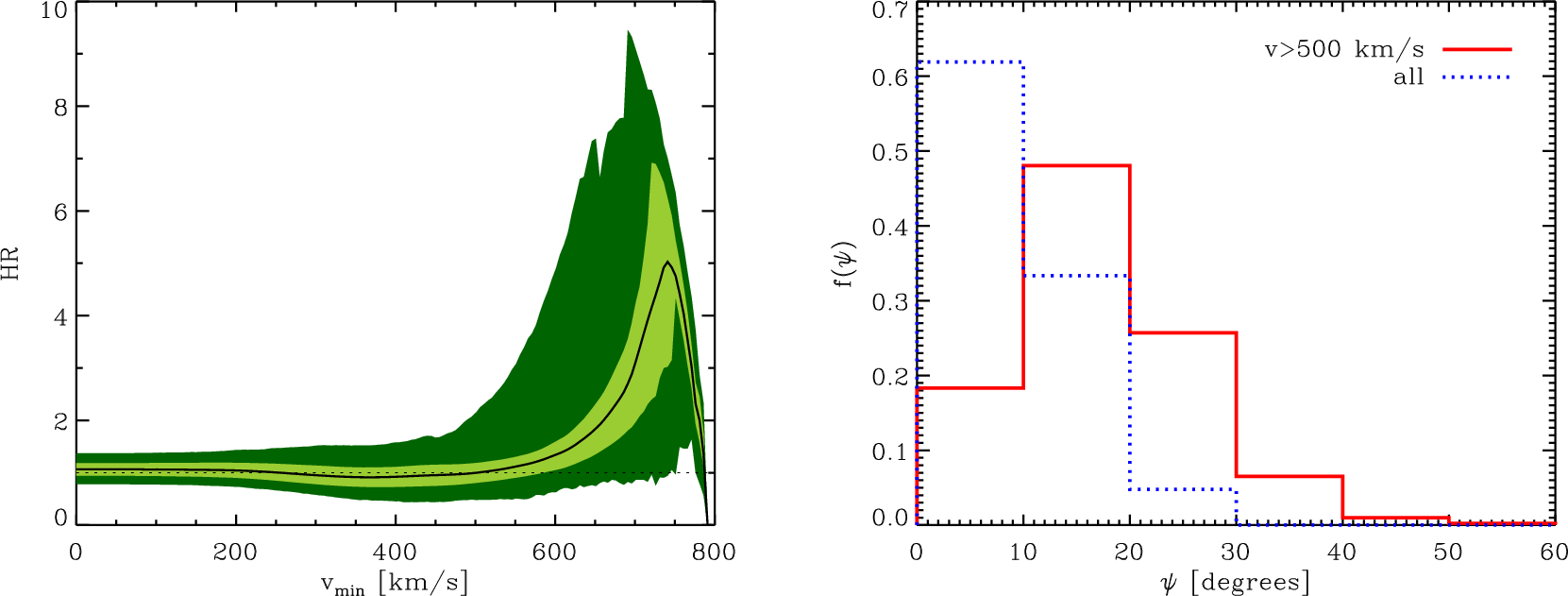}
  \caption{The distribution of HR$(\vmin)$ as a function of $\vmin$
    (\textit{left}) and of $\psi$ for $\vmin=0$ and 500 km/s
    (\textit{right}) for the VL2 simulation.}
\label{fig:HR+psi}}

This is just one strong example, and we would like to understand what
sorts of hotspots we might expect. For this purpose we define the
hotspot ratio
\begin{equation}
{\rm HR}(\vmin) = \frac{{\rm max}\left\{p_i(\vmin)\right\}}{{\rm max}\left\{p^{\rm MB}_i(\vmin)\right\}},
\end{equation}
the ratio of the hottest pixel in the sphere sample skymap above
$\vmin$ to the hottest pixel in the corresponding MB-case, and the
hotspot angle $\psi$ between these two pixels. We calculate HR and
$\psi$ for all 100 sample spheres, and in each sphere for a full
$2\pi$ rotation (in one degree increments) of the direction of Earth's
motion. We show in Fig.~\ref{fig:HR+psi} for VL2 the distribution of
HR as a function of $\vmin$ and the distribution of $\psi$ for the
case without a velocity threshold and for $\vmin=500$ km/s. For small
velocities the mean of HR is unity and the r.m.s. variation is only
10\%. As $\vmin$ increases, HR grows: at $\vmin=600$ km/s, the mean HR
is $1.3 \pm 0.35$, and at $\vmin=700$ km/s it's $3.1 \pm 1.4$. The
downturn at the very highest velocities is caused by running out of
particles in the sample spheres. There are also marked changes in the
direction of the hottest pixel. Even without a velocity threshold
($\vmin=0$ km/s) in 38\% of all cases the direction of the hottest
spot on the sky is more than 10$^\circ$ removed from the direction of
Earth's motion (i.e. the MB halo expectation). \textit{At} $\vmin=500$
km/s \textit{the bulk of DM particles are more likely to be coming
  from a local hotspot with $\psi>10^\circ$ than from the direction of
  Earth's motion.} In this case there is only an 18\% chance of having
$\psi < 10^\circ$. Not all of this structure is due to individual
subhalos; some of it may arise from tidal streams, and some from the
anisotropic velocity distribution of the host halo particles.

\section{Implications for DAMA}
\label{sec:dm_models}
In light of the DAMA/LIBRA results \cite{Bernabei2008}, two scenarios
which have garnered a great deal of attention of late are light dark
matter \cite{Bottino:2003cz,Gondolo:2005hh}, and inelastic dark matter
\cite{Smith:2001hy, Tucker-Smith:2004jv}. These scenarios were
proposed some time ago to address the conflicts between DAMA
\cite{Bernabei:2000qi} and CDMS \cite{Abusaidi:2000wg} as well as
other experiments at the time.

Recent studies
\cite{Chang:2008gd,March-Russell2009,Cui:2009xq,Arina:2009um} have
shown that iDM remains a viable explanation of the DAMA data,
consistent with recent results from CDMS \cite{Ahmed:2008eu}, XENON10
\cite{Angle:2007uj}, KIMS \cite{Lee.:2007qn}, ZEPLIN II
\cite{Alner:2007ja}, ZEPLIN III \cite{Lebedenko:2008gb} and CRESST
\cite{Angloher:2008jj}. Such models have some tension with CRESST (see
the discussion in \cite{Chang:2008gd,SchmidtHoberg:2009gn}), which
observed 7 events in the tungsten band, while approximately 13 would
be expected \cite{Chang:2008gd}. While lighter masses have no tension
with CDMS, higher mass ($\gtrsim 250$ GeV) WIMPs do.

Light dark matter no longer works in its original incarnation
\cite{Bottino:2003cz,Gondolo:2005hh}, but instead relies upon
``channeling,'' \cite{Bernabei:2007hw,Petriello:2008jj}, a difficult
to quantify effect whereby some fraction of nuclear recoils have
essentially all of the energy converted into scintillation light,
rather than just a fraction. Even including channeling, such scenarios
seem to have strong tension with the data
\cite{Chang:2008xa,Fairbairn2009,Savage:2008er}, both in the spectrum
of the modulation, and constraints from the unmodulated event rate at
low (1-2 keVee) energies\footnote{keVee stands for ``electron
  equivalent keV'', a unit of energy used for scintillation light,
  which is produced by interactions of the recoiling nucleus with
  electrons. It is related to the full nuclear recoil energy (in keVr)
  through the quenching factor $q$: $E({\rm scintillation})/{\rm
    keVee} = q\;E({\rm recoil})/{\rm keVr}$. $q$ is a
  material-dependent quantity that must be experimentally
  determined.}. If one disregards the lowest (2-2.5 keVee) bin from
DAMA, however, the fit improves \cite{Chang:2008xa,Hooper:2008cf}, but
there is still tension for the lightest particles from the presence of
modulation above 4 keVee and an overprediction of an unmodulated
signal at lower energies. It is important to note that the
uncertainties in $L_{\rm eff}$, the scintillation efficiency of liquid Xenon (see \cite{Manzur:2009hp} for a discussion),
are not completely included in these analyses, and the
most recent measurements of $L_{\rm eff}$ \cite{Manzur:2009hp} suggest
that the low energy analysis threshold may be somewhat higher than the
4.5keVr used with $L_{\rm eff}=0.19$, weakening the limits. On the
other hand, the most recent analysis from XENON10
\cite{Collaboration:2009xb} using a more advanced rejection of
double-scatter events shows no events at all in the signal region all
the way down to the S2 threshold, strengthening the limits. In light
of these uncertainties, and in view of the importance of the result,
we believe it is best to maintain an open mind about the viability of
the light WIMP explanation.

For our purposes, the crucial feature is that these scenarios sample
the high ($v \sim 600$ km/s) component of the velocity distribution,
and so are especially sensitive to the departures from a simple
Maxwell-Boltzmann distribution. Consequently, it is important to
investigate whether these changes affect the tensions described above.

\subsection{Inelastic Dark Matter}
The inelastic Dark Matter scenario \cite{Smith:2001hy} was proposed to
explain the origin of the DAMA annual modulation signal. The scenario
relies upon three basic elements: first, the presence of an excited
state $\chi^*$ of the dark matter $\chi$, with a splitting $\delta =
m_{\chi^*}-m_\chi \sim m_\chi v^2$ comparable to the kinetic energy of
the WIMP. Second, the absence of, or at most a suppressed elastic
scattering cross section off of nuclei, i.e., the process $\chi N
\rightarrow \chi N$ should be small. Third, an allowed cross section
for inelastic scatterings, i.e., $\chi N \rightarrow \chi^* N$, with a
size set roughly by the weak scale.

Although these properties may seem odd at first blush, they are in
fact perfectly natural if the scattering occurs through a gauge
interaction \cite{Smith:2001hy,Tucker-Smith:2004jv}, where the
splitting is between the two Majorana components of a massive
pseudo-Dirac fermion, or between the real and imaginary components of
a complex scalar. Simple models can be constructed where the mediating
interaction is the Z-boson
\cite{Smith:2001hy,Tucker-Smith:2004jv,Arina:2007tm,Cui:2009xq}. Models
with new vector interactions to explain the PAMELA positron excess
also naturally provide models of iDM \cite{ArkaniHamed:2008qn}, and
often where all scales arise naturally from radiative corrections
\cite{ArkaniHamed:2008qn,ArkaniHamed:2008qp,Baumgart:2009tn,Cheung:2009qd,Katz:2009qq}. Composite models provide a simple origin for the excited states as well \cite{Alves:2009nf, Lisanti:2009am}. 

The principle change is a kinematical requirement on the scattering,
\begin{eqnarray}
\beta_{\rm min} &=& \sqrt{\frac{1}{2 m_N E_R}} \left( \frac{m_N E_R}{\mu} + \delta \right),
\label{eq:betamin}
\end{eqnarray}
where $m_N$ is the mass of the target nucleus and $\mu$ is the reduced
mass of the WIMP/target nucleus system.  For $\delta \sim \mu
\beta^2/2$, the consequences can be significant. This requirement has
three principal effects: first, the kinematical constraint depends on
the target nucleus mass, and is more stringent for lighter
nuclei. If we are sampling dominantly the tail of the velocity
distribution, the relative effect between heavy and light targets
(e.g., iodine versus germanium) can be significant. Second, again
because the signal is sampling the tail of the velocity distribution,
the modulated amplitude can be significantly higher than the few
percent expected for a standard WIMP. In fact, in the cases where
there are particles kinematically scattering in the summer, but not
the winter, the modulation can reach 100\%. Third, the inelasticity
suppresses or eliminates the event rate at lower energies. Because
standard WIMPs have exponentially more events at lower energies, most
experiments have focused on controlling background in this region and
lowering the threshold. In particular, the XENON10 and ZEPLINIII
experiments have few events at low energies, but a significant
background at higher energies. Consequently, their limits for standard
WIMPs are quite strong, but significantly weaker for inelastic WIMPs.
The combination of these three elements allows an explanation of the
DAMA result that is consistent with all other current experiments
\cite{Chang:2008gd,March-Russell2009,Cui:2009xq}.

\subsubsection{Fitting the DAMA signal}
DAMA reports an annual modulation in the range of $2-6$ keVee
range. This can be interpreted as arising from either Na or I
scattering events. In the former case (Na), it has been shown that the
``light inelastic'' region (an approximately 15 GeV WIMP with $\delta
\approx 30 \keV$) can open up significant parameter space
\cite{Chang:2008xa,SchmidtHoberg:2009gn}. In the latter case (I), we
find the ``heavy inelastic'' region (a 100+ GeV WIMP with $\delta
\gsim 100 \keV$) opens significant parameter space. Since the
constraints are stronger on the heavy case, we focus on this region.

When determining the precise values of parameters that might agree
with DAMA, we must convert from keVee to keVr, which already
introduces significant uncertainties, a point which has been recently
discussed by \cite{March-Russell2009}. The quenching factor for iodine
has been found to have a range of different values, including $q=0.09
\pm 0.01$ \cite{Bernabei:1996vj}, $q=0.05 \pm 0.02 $
\cite{Fushimi:1993nq}, $q=0.08 \pm 0.02$ \cite{Gerbier:1998dm} and
$q=0.08 \pm 0.01$ \cite{Tovey:1998ex} \footnote{The value quoted by
  \cite{Tovey:1998ex} is generally $q=0.086 \pm 0.007$, but this value
  averages over a wide range of energies. The two measured values for the
  DAMA energy range specifically are $q=0.08 \pm 0.01$ and $q=0.08 \pm
  0.02$.}. The first two measurements of the quenching factor are
somewhat indirect, fitting event distributions at low energies. The
last two are more direct. Of the last two, we should note that the
first includes non-linearity in its stated uncertainty, while the
second explicitly fixes to a signal electron energy, and thus this
effect is not included. We adopt $q=0.08 \pm 0.02$ as the value for
quenching, which corresponds to a range of $25-75$
keVr for DAMA. 

\FIGURE[t]{
  \includegraphics[width=\textwidth]{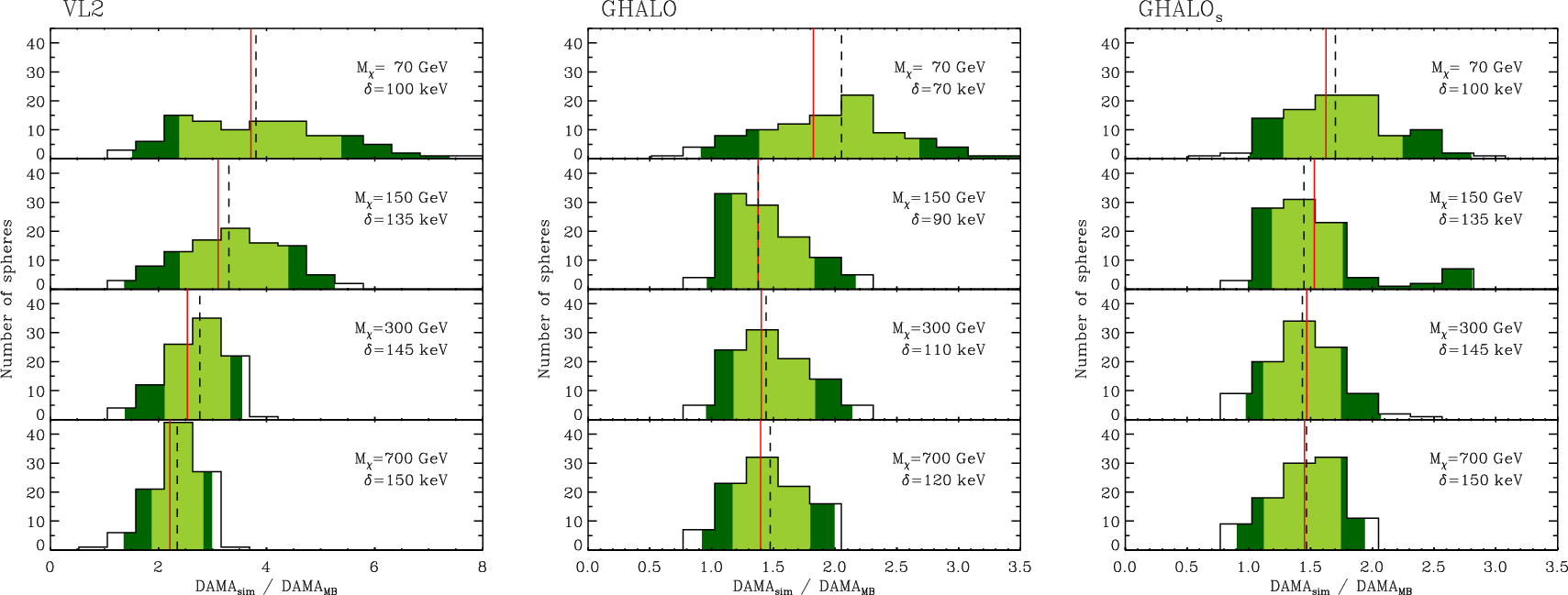}
  \caption{The ratio of DAMA signal in a given simulation to that in a
    MB distribution. The solid red line is the spherical shell
    average, the dashed line the median of the distribution over the
    sample spheres.}
\label{fig:damaratio}}

\subsubsection{Constraining the iDM interpretation of DAMA}
While iDM allows DAMA to exist consistently with the other
experiments, the reach of other experiments is at or near the
predicted DAMA level. In particular, as we see in
Fig.~\ref{fig:contour700}, within the context of a MB halo, at high
masses, CDMS excludes the entire range of the parameter space that
fits DAMA. At the same time, the CRESST results (with a tungsten
target) seem very tense with the data over the whole mass range. These
two experiments provide the greatest present tension with the iDM
interpretation of DAMA. An immediate question we can ask is then
whether the velocity-space distribution of particles in the
simulations makes the constraints more or less significant than a MB
distribution.

This is not an easy question to answer. Since there can be significant
spatial variation in the velocity distribution, we would like to
quantify this without making full exclusion plots for every data
point.  The three principle constraints on the iDM interpretation of
DAMA come from a) Xe experiments (XENON10, ZEPLIN-II and ZEPLIN-III),
b) CDMS (Germanium) and c) CRESST (Tungsten). While varying the halo
model can have significant effects on each experiment, including rates
and spectra, the variation of velocity structures in the different
halos affects these limits differently, and it is difficult to
quantify in aggregate what the effects are.

Before proceeding into a detailed analysis of the DAMA signal, we can
already study a preliminary question: how does the cross section
needed to explain DAMA compare between a MB halo and a simulation? We
consider the ratio of the modulation between MB and the simulations in
Fig.~\ref{fig:damaratio}. Overall, we see that the signal at DAMA is
\textit{increased} in the simulation, as much as a factor of a few.

Of course, increasing the DAMA signal does not change the constraints
if the signal in any of the other experiments is changed, as well. We
proceed by next considering a ratio of ratios (RoR). Since Germanium
(CDMS) has a higher velocity threshold than Iodine (DAMA), while
Tungsten (CRESST) has a lower velocity threshold, a simple test is to
look at the variation of the signal at different experiments as a
fraction of the DAMA modulation amplitude. We thus calculate for each
of the simulation samples (spherical shell and 100 spheres), as well
as for the best-fitting MB model, the ratios of the CDMS and CRESST
signals, integrated from 10-100 keV, to the total DAMA modulation in
either of the high-$q$ and low-$q$ benchmark regions described
above. We then divide the simulation ratio by the MB ratio:
\mbox{(CDMS/[DAMA])$_{\rm sim}$/(CDMS/[DAMA])$_{\rm MB}$} and
\mbox{(CRESST/[DAMA])$_{\rm sim}$/(CRESST/[DAMA])$_{\rm MB}$}. If
these RoR's are smaller than one, it suggests that using the
simulation's velocity distribution instead of the best-fitting MB
distribution weakens the limits relative to DAMA, while values larger
than one imply stronger limits. The effects on Xenon limits are harder
to quantify, because of the added uncertainties in the conversion from
keVee to keVr at those experiments, and where, precisely, the
backgrounds lie. Thus, we consider first only CDMS and CRESST limits.

\FIGURE[t]{
  \includegraphics[width=\textwidth]{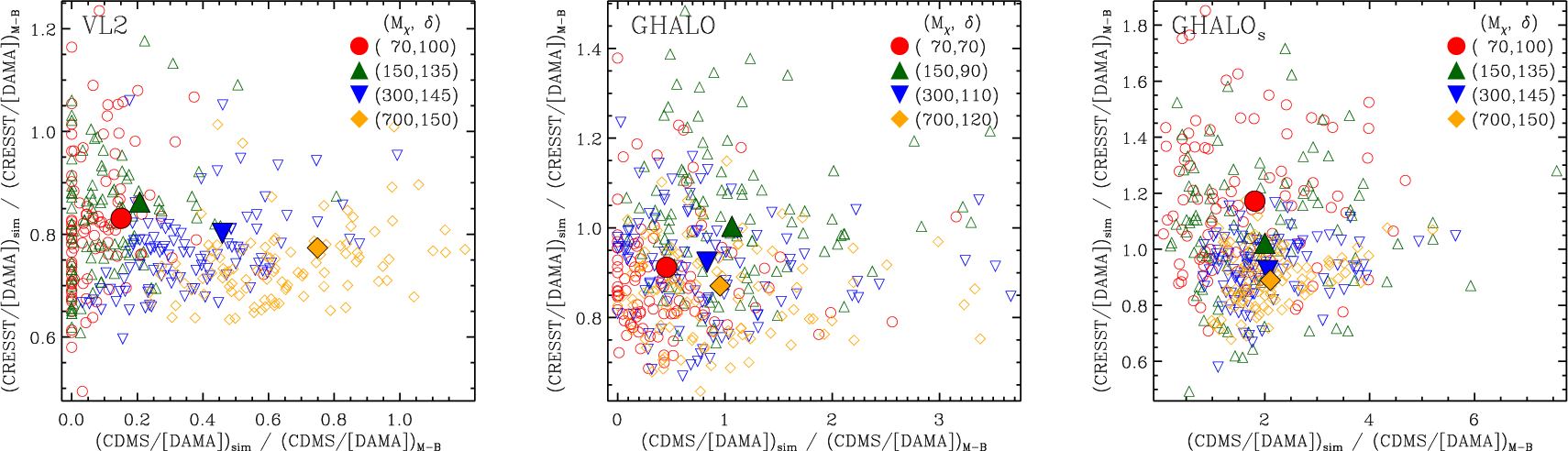}
  \caption{Scatter plots of the ratios of ratios (RoR, see text for
    details) for the CRESST and CDMS experiments. RoR's less than one
    indicate that using the simulation's velocity distribution instead
    of the best-fitting MB model weakens the CRESST or CDMS limits
    relative to the DAMA (high-$q$) signal. 
}
\label{fig:ratorathighq}}

\FIGURE[t]{
  \includegraphics[width=\textwidth]{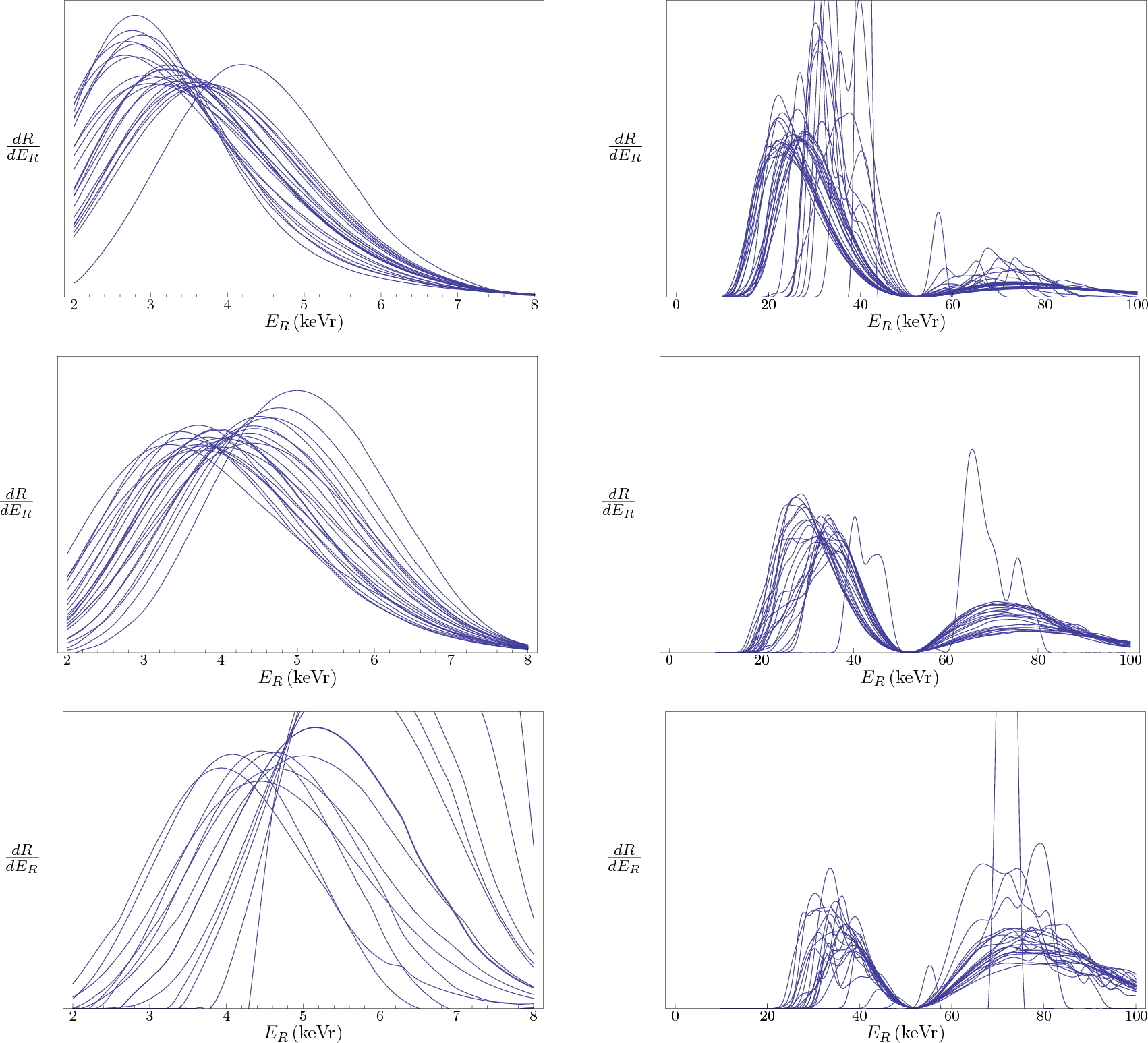}
  \caption{\textit{Left:} The differential recoil energy distribution
    (in dru$=$events/kg/day/keV) at DAMA for selected spheres from
    VL2, GHALO and \ghalos, normalized to unity in
    the DAMA signal range 2-6 keVee \textit{Right:} The distribution
    at CRESST for the same spheres, normalized to unity from 0-100
    keV. The models are $m_\chi = 100 \gev$ and $\delta = 130\, ({\rm top}), 150\, ({\rm middle}),
    170\, ({\rm bottom}) \kev$.}
\label{fig:spectra100}}

\FIGURE[t]{
  \includegraphics[width=\textwidth]{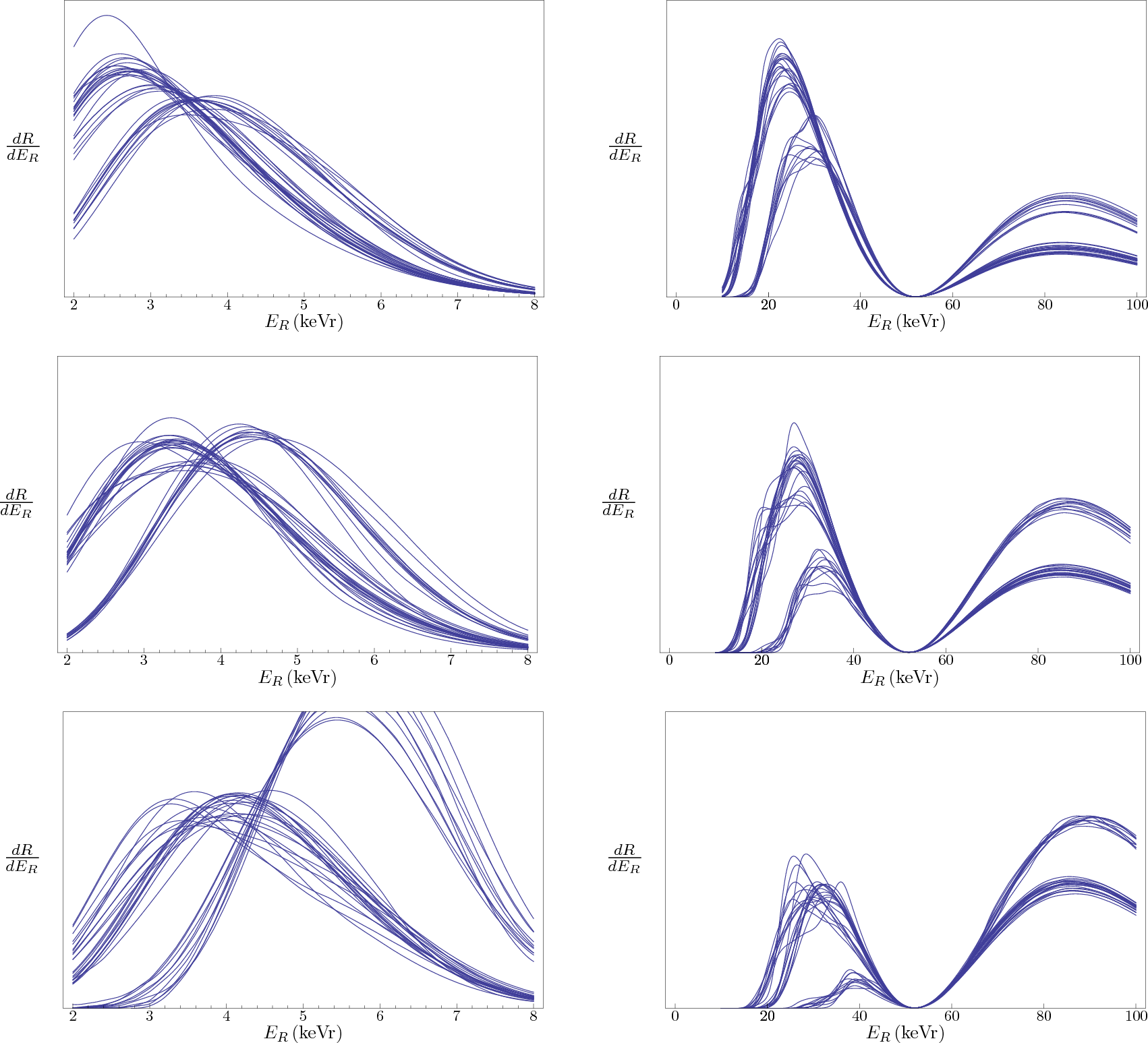}
  \caption{As in Fig.~\ref{fig:spectra100}, but with
    $m_\chi=700\gev$.}
\label{fig:spectra700}}

In Fig.~\ref{fig:ratorathighq} we show scatter plots of the CRESST RoR
against the CDMS RoR. Large filled symbols indicate the spherical
shell sample and small symbols are used for the sample spheres. Note
that in some cases the CDMS RoR is zero, indicating that the
simulation velocity distribution resulted in no CDMS signal at all. A
few things are immediately obvious from this plot. First, the limits
from CDMS can vary wildly between simulations, and even between
different spheres within a single simulation. This simply represents
how dramatically the velocity distribution can change at the highest
velocities. Second, we see that the CRESST rate is much more weakly
affected, with typically suppressions of (0.6-1), (0.7-1.1), and
(0.7-1.3) for Via Lactea II, GHALO and \ghalos\, respectively. Thus,
from the perspective of Poisson limits, these results would suggest
that those derived from Maxwellian halos are possibly excessively
aggressive by almost a factor of two.  However, many limits are placed
using one of Yellin's techniques \cite{Yellin:2002xd}, where not only
the overall rate, but also the distribution of signal versus
background is important. Here we find that the halo uncertainties can
be at their largest.  We show in Fig.~\ref{fig:spectra100} the spectra
at DAMA and CRESST for a 100 GeV WIMP with $\delta=130,150,170
\kev$. We employ energy smearing at DAMA by assuming that the smearing
reported for the one of 25 targets of DAMA/LIBRA
\cite{Bernabei:2008yh} is characteristic of all of them. We assume a
smearing of 1 keV at CRESST.

One can see that the peak positions and properties can change by quite
a large amount. At DAMA, the effect of this is principally to shift
the peak. Such an effect is largely degenerate with the quenching
uncertainty, which can reasonably range from $q=0.06$ to
$q=0.09$. (For a lower quenching value, for instance, the peak will
shift to lower energy in keVee.) The effects would be similar at
XENON10, where the energy smearing will eliminate most interesting
structures, and the shift in the location of the peak will be
comparable to the uncertainties induced from $L_{\rm eff}$. In
contrast, the spectrum for CRESST can vary \textit{dramatically}, with
the peak location moving from below $30 \kev$ to above $60 \kev$. The
implication of this is that techniques such as optimum interval,
maximum gap and ${\rm p_{max}}$ are likely all overly aggressive in
that a peak in the spectrum might arise at a specific location in the
real Milky Way halo, but that would not be reproduced at the
appropriate position, for instance in a Maxwellian halo. In general, any technique that relies on knowing precisely the predicted spectrum is unreliable when considering these variations.

\FIGURE[t]{
  \includegraphics[width=\textwidth]{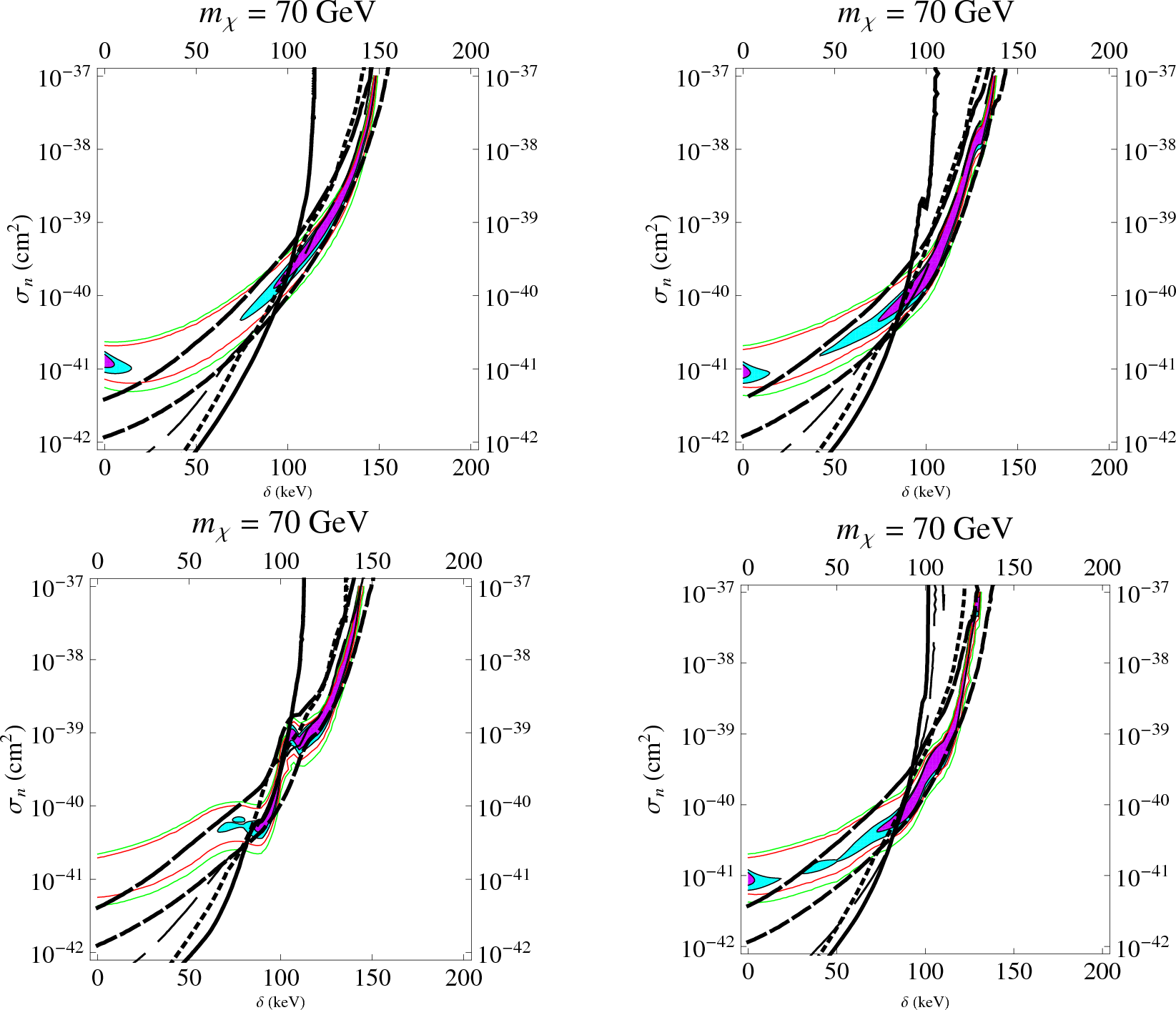}
  \caption{Allowed parameter space for DAMA at 90\% (purple) and 99\%
    (blue) with $m_\chi=70\gev$. For comparison the 90\% and 99\%
    regions for DAMA/LIBRA rates only (as opposed to DAMA/NaI +
    DAMA/LIBRA) are shown in red and green lines. Constraints are CDMS
    (solid), ZEPLIN-III (long dashed, thin), ZEPLIN-II (long-dashed,
    thick), CRESST (medium-dashed) and XENON10
    (short-dashed). The regions are MB with $v_0=220\kms$ and $v_{\rm
      esc}=550\kms$ {(\em top left)}, VL2 {(\em top right)}, and a
    sample sphere from VL2 \textit{(bottom, left)} and
    \ghalos\ \textit{(bottom, right)}.}
\label{fig:contour70}}

\FIGURE[t]{
  \includegraphics[width=\textwidth]{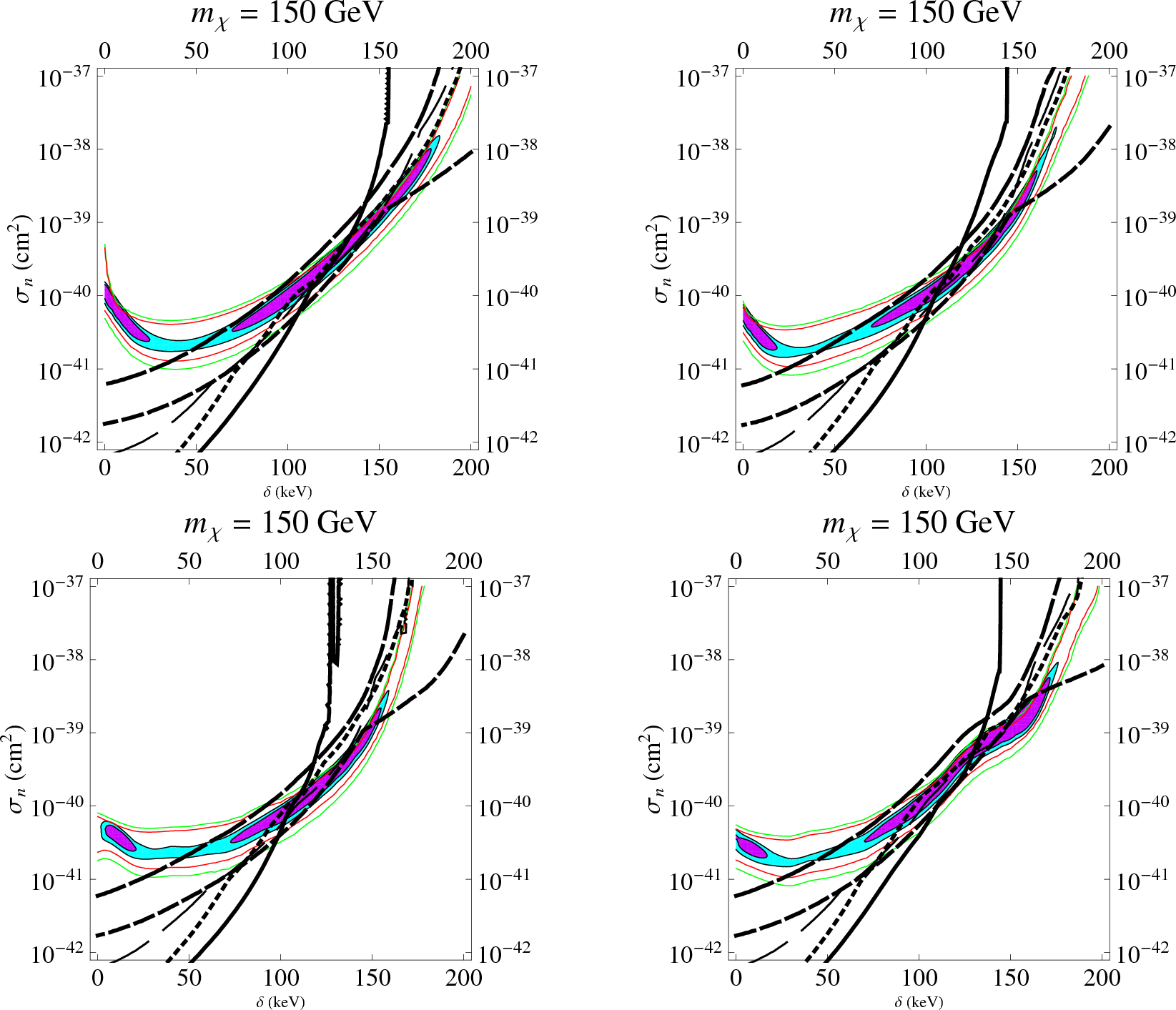}
  \caption{As in \ref{fig:contour70}, but with $m_\chi = 150\gev$.}
\label{fig:contour150}}

\FIGURE[t]{
  \includegraphics[width=\textwidth]{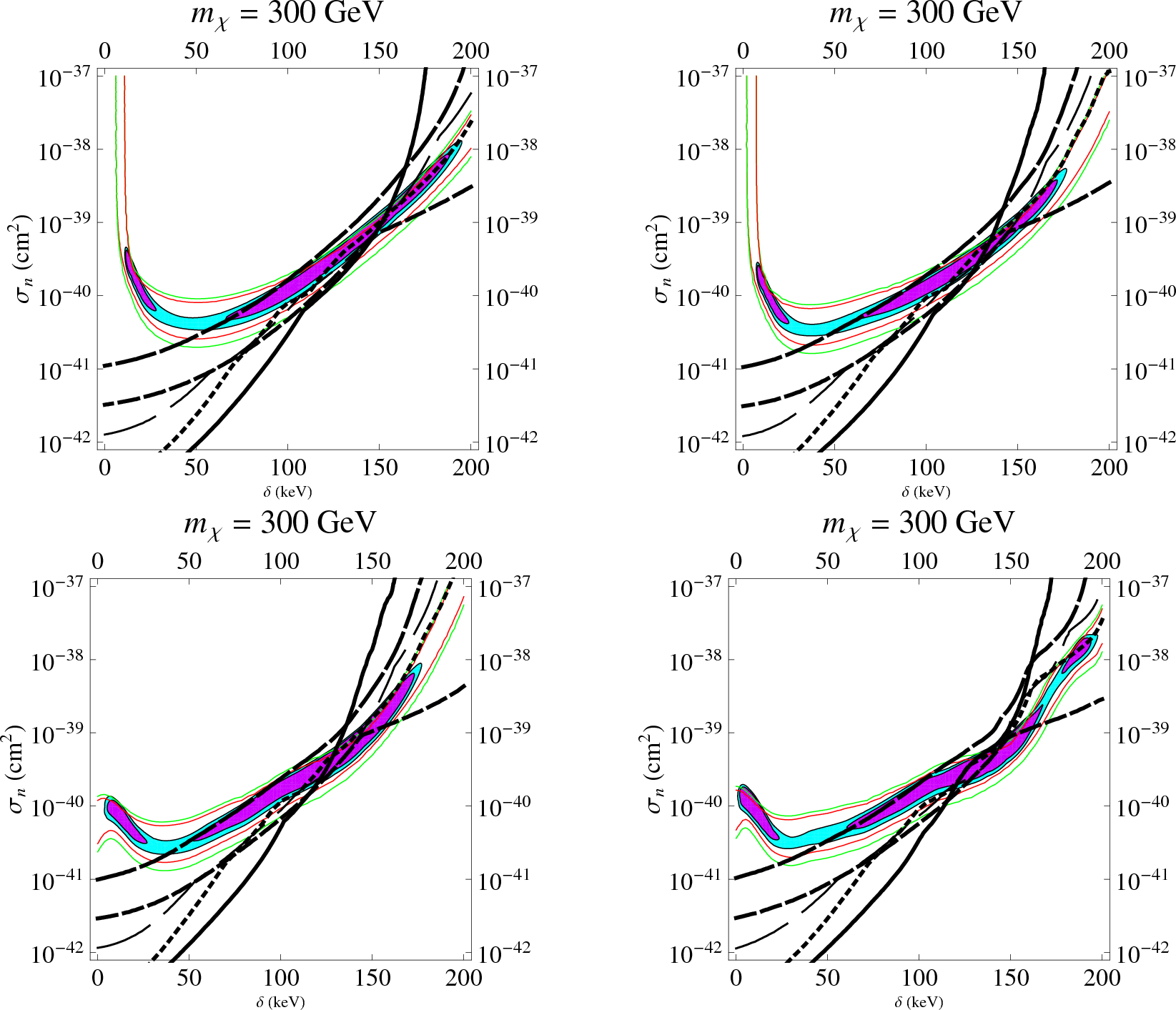}
 \caption{As in \ref{fig:contour70}, but with $m_\chi = 300\gev$.}
\label{fig:contour300}}

\FIGURE[t]{
  \includegraphics[width=\textwidth]{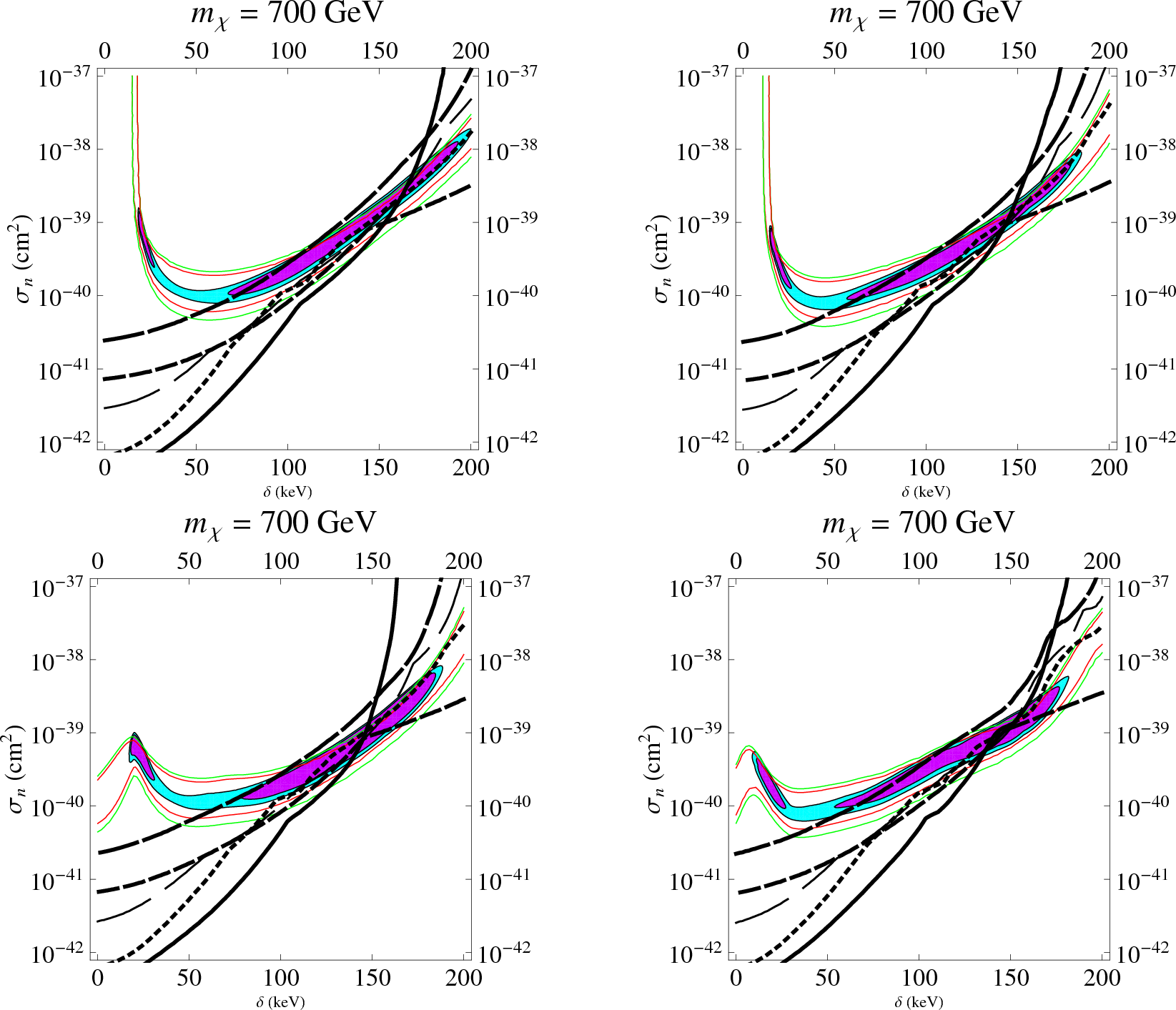}
 \caption{As in \ref{fig:contour70}, but with $m_\chi = 700\gev$.}
\label{fig:contour700}}

Although we are still limited by our sample of simulations, we can
study the effects on various limits by looking at the allowed
parameter space and limits in a variety of halos. To do so, we largely
follow \cite{Chang:2008gd} in calculating the allowed parameter space
and limits, with a few important differences. As pointed out by
\cite{March-Russell2009}, the uncertainty in quenching factor for
iodine can be extremely important. We thus consider the range $q=0.06$
to $q=0.09$ and take the 90\% allowed region to be the union of 90\%
allowed regions, and similarly for the 99\% region. We smear the
signal as described above.

For CDMS we employ the maximum gap method with the data set specified in \cite{Chang:2008gd}.

For limits arising from XENON10, we use the data presented in the
recent reanalysis of \cite{Collaboration:2009xb}, and calculate the
maximum gap limit (as in ZEPLIN-II and ZEPLIN-III) for both \cite{Sorensen:2008ec} and
\cite{Manzur:2009hp} values of $L_{\rm eff}$, taking at every point the
more conservative of the two. For limits, we employ the maximum gap
method. Since the data are smeared and there are great uncertainties
in $L_{\rm eff}$, the detailed spectral information arising from structure
is lost.

For ZEPLIN-II, we take the data described in \cite{Alner:2007ja}, but
(conservatively) take the energy values to be shifted by a factor of
$0.24/0.19$ for a maximum gap analysis, which employs the value of $L_{\rm eff}$ from
\cite{Sorensen:2008ec} at higher energies.\footnote{We could
  parameterize $L_{\rm eff}$ as a function of energy, and use
  different shifts at different energies, but the limits are
  essentially the same, being dominated by the events at the highest
  energy.}

For ZEPLIN-III, we utilize the data within the published acceptance
box up to 15 keVee for the maximum gap analysis. We additionally employ data outside the blind
acceptance box up to 30 keVee, the maximum to which efficiencies were
provided in \cite{Lebedenko:2008gb}. We use the delineated $1\sigma$
region for the data in this high energy range since the $3\sigma$ region is not specified. To convert from
keVee to keVr, we use
\begin{equation}
E_r = (0.142 E_{\rm ee} + 0.005) \, \exp(-0.305\,E_{\rm ee}^{0.564})
\end{equation}
below 10 keVee \cite{March-Russell2009,Arina:2009um}, and a constant
ratio of 0.55 above that. (We use a larger value than
\cite{Arina:2009um} at high energies to account for the energy
dependent value of $L_{\rm eff}$ suggested by \cite{Sorensen:2008ec},
which tends to weaken limits slightly.)  Note that we do not employ a
background subtraction using the science data as in
\cite{Cline:2009xd}, nor do we do so for ZEPLIN-II as the same basic
technique seems to have led to a significant overprediction of
background for ZEPLIN-III.

For all Xe based experiments, we take a fixed smearing of $\sqrt{30}$
keVr, which is typical of the smearing of the experiments in the range
of the peak signal.

For the CRESST experiment, we consider the data from the commissioning
run \cite{Angloher:2008jj}, as well as the additional events at $\sim$
22\kev, 33\kev\, and 88\kev\, presented for Run 31 by \cite{probst}, and
take a smearing of 1 keV. As described above, Yellin-style techniques
are ironically inappropriate for experiments with high energy
resolution. Because of the significant variation in the possible
spectrum when varying halo models with the high energy resolution, we use a binned Poisson analysis,
rather than a maximum-gap technique. We divide up the signal region
into a low region ($E_R< 35 \kev$), a middle region ($35 \kev < E_R <
50 \kev$) and a high region $(50 \kev < E_R < 100 \kev)$. In this way,
we do not overweight the events near the zero of the form factor (in
the mid region), but still include some broad spectral information. We
require that the rate be lower than 95\% Poisson upper limits for each
bin, which is a 95\% limit when the signal is exclusively in the low
bin and $\sim$ 90\% when there is signal in both the low and high
bins, and $\sim$ 85\% in the (rarer) instances when there is signal in
all three bins. Although stronger limits can be set using for
instance, optimum interval, because of the uncertainties in the halo
distribution, this seems inappropriate. In other words, if some of the
events at CRESST are real, then the distribution of events may well be
telling us the properties of the halo.

We show the results of these limits in
Figs.~\ref{fig:contour70}-\ref{fig:contour700}. Even when accounting
for smearing, and using a binned Poisson limit instead of optimum
interval, it is remarkable how significantly the allowed parameter
space can change from simulation to simulation, and between spheres in
the same simulation. We see that CRESST remains the most constraining,
as found previously, but we see important differences.  First,
agreement is generally improved in these spheres. This is due to the
presence of structure at high velocities which increases the modulated
fraction at DAMA. As a consequence, small regions of allowed parameter
space exist at high mass, the size of which depends significantly on
the particular halo. Larger regions exist at smaller mass ($\sim 150 \gev$). 
Moreover, the allowed region of parameter space can shift
dramatically in $\sigma$ and $\delta$, with pockets appearing at large
$\delta$ from high velocity particles. Should iDM be relevant for
nature, the detailed nature of the halo is clearly significant in
determining the properties of the direct detection signals.

\FIGURE[t]{
  \includegraphics[width=\textwidth]{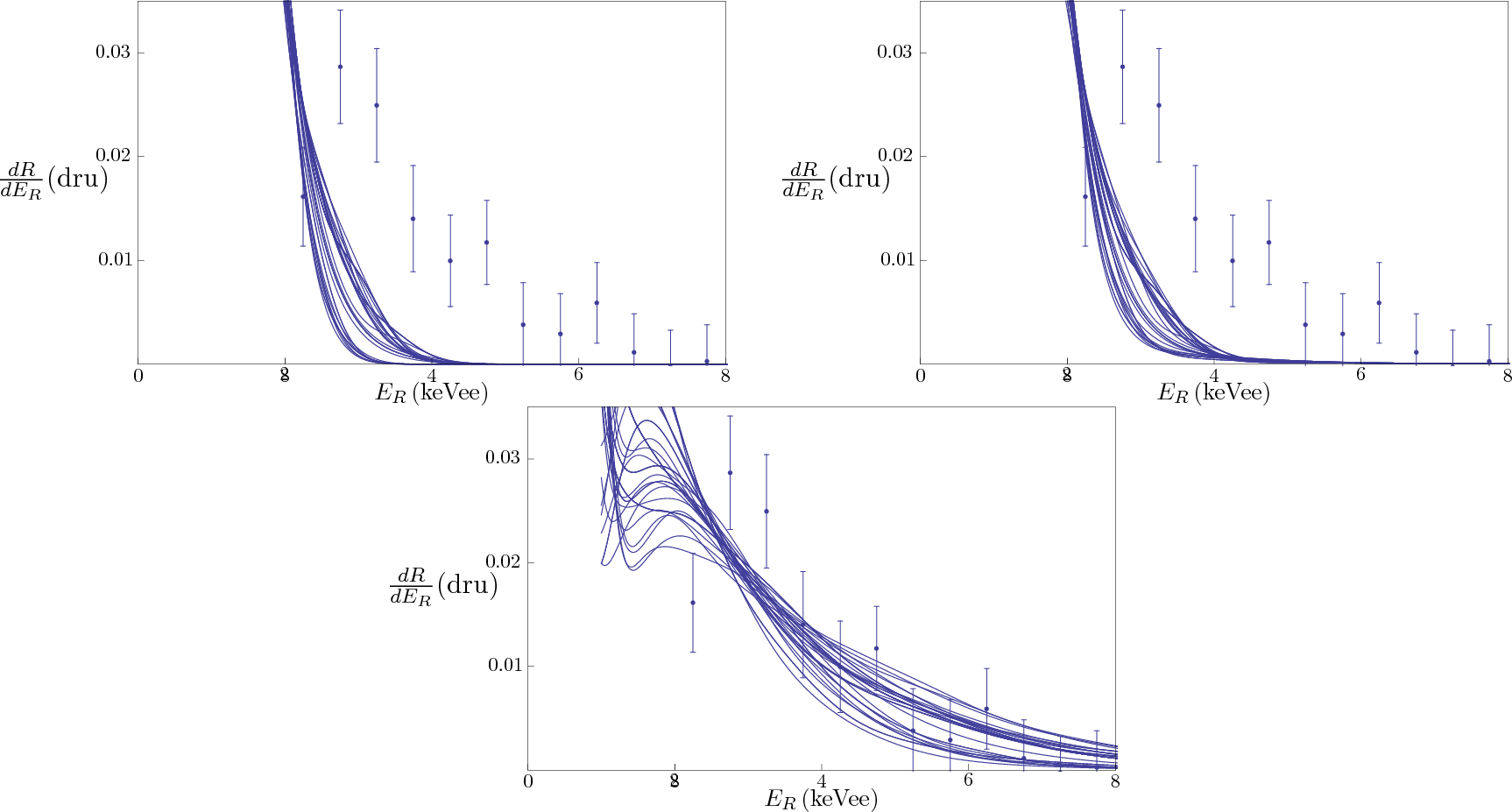}
  \caption{Spectra for a sample of spheres best fit to the DAMA data (shown) within the different
    simulations for LDM models with $m_\chi = 3 \gev$ (\textit{top,
      left}), $7\gev$ (\textit{top, right}), and $13\gev$ ({\em
      bottom}).}
\label{fig:LDMspec}}

\FIGURE[t]{
  \includegraphics[width=0.9\textwidth]{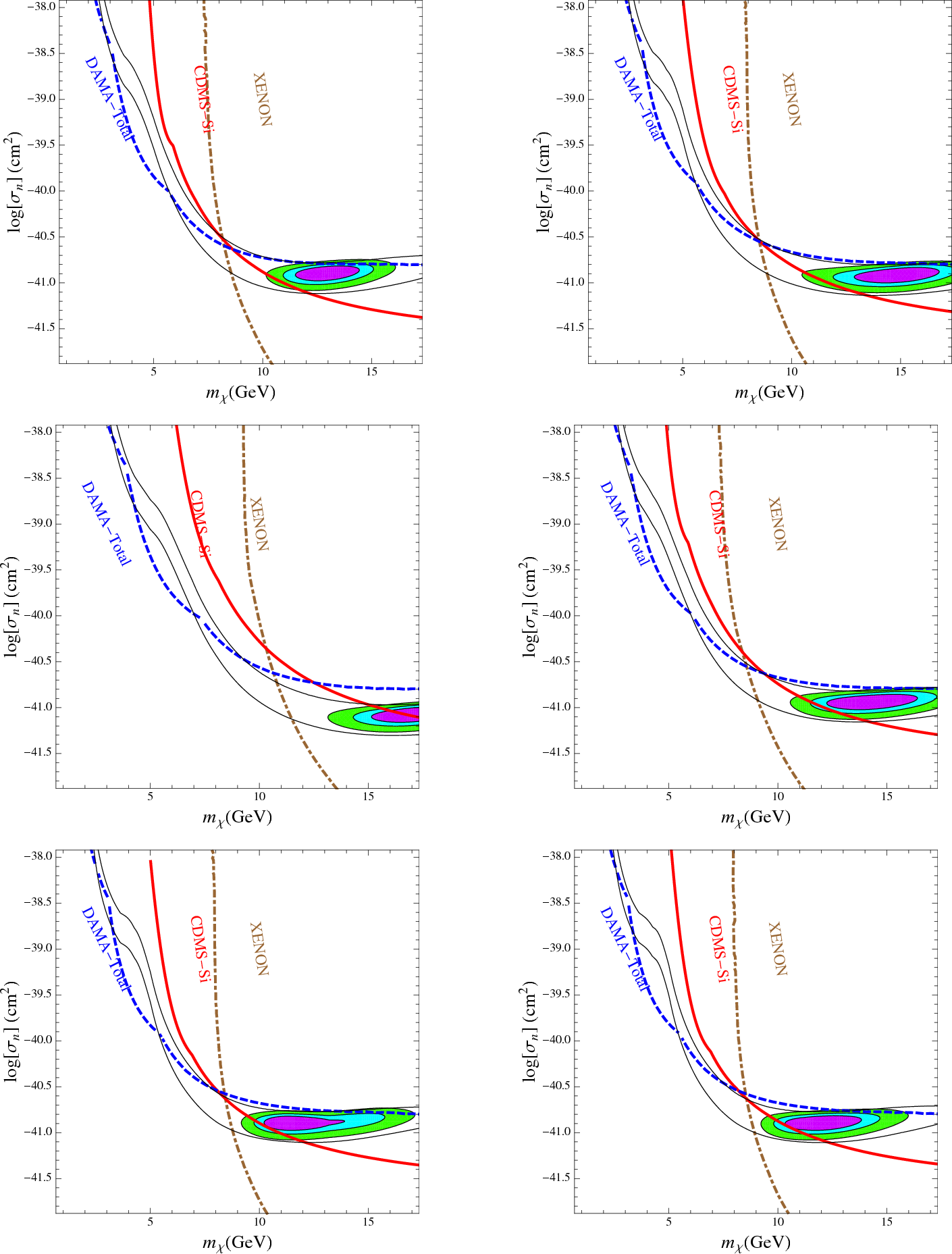}
  \caption{The allowed parameter space for LDM models. \textit{Top,
      left:} MB with $v_0=220\kms$ and $v_{\rm esc}=550\kms$. \textit{Top,
      right:} VL2. \textit{Middle, left:} GHALO. \textit{Middle, right}
    \ghalos. \textit{Bottom} two sample spheres taken from the
    simulations.}
\label{fig:paramlight}}

\subsection{Light dark matter}
Light ($\sim$ GeV) dark matter \cite{Bottino:2003cz} has been proposed
as a means to reconcile DAMA with the other existing experiments
\cite{Gondolo:2005hh, Bottino:2008mf, Petriello:2008jj}. The current
iteration of LDM involves ``channelling'' \cite{Bottino:2007qg,
  Bernabei:2007hw} whereby for some small fraction of the time the
entirety of the nuclear recoil energy is converted to scintillation
light. Since the observed energy is lower, it allows lighter WIMPs to
scatter at the DAMA experiment. Such light WIMPs may be incapable of
depositing energy above the threshold at XENON10 or CDMS-Si, for
instance, allowing DAMA to evade these bounds.

Such an explanation is not without difficulty, however. These light
WIMPs have a rapidly falling event spectrum, due to the decreasing
probability of channeling at higher energies, and the increasingly
suppressed number of particles with higher kinetic energy. As a
consequence the spectrum of LDM seems generally to fall too rapidly at
light masses \cite{Chang:2008xa,Fairbairn2009,Savage:2008er}. At
higher masses, the spectrum is acceptable, but is excluded by other
experiments. We should note again here that this is under the
assumption that the errors in the data are statistical only. Should
there be, e.g., significant changes in the efficiency, it is possible
that LDM could give a better fit. As noted earlier, the
uncertainties in $L_{\rm eff}$ should weaken these limits, while the most
recent analysis from XENON10 (with no events to the S2 threshold)
would strengthen the limits.  Thus, while the models do not seem to
describe the data well, there are adequate uncertainties that this
scenario remains an interesting possibility as an explanation of the
DAMA signal.

Since these models are also extremely sensitive to the presence of
high velocity particles, one might wonder whether, just as in the iDM
case, the properties of halos at high velocities might modify the
spectrum and the relative signal strength at other experiments.

We begin again by addressing the question of the effects on the
spectrum. We show these for three different masses in
Fig.~\ref{fig:LDMspec}. One can see that, while there are noticeable
changes in the spectrum, its overall shape remains basically
unchanged. The difference between LDM and iDM here is simple: in iDM,
the ``threshold'' (i.e., minimum velocity) scattering is actually at
large energy, and to go to lower energies one requires \textit{higher}
velocities. Competing against this are the significant form factor
effects. The competition between these two sensitive quantities leads
to dramatic changes and features. For LDM, the threshold scattering is
at zero, and all effects (channeling, form factor, velocity
distribution) contribute to the falling (unmodulated) spectrum. Thus,
it can fall somewhat more or less rapidly, depending on the number of
particles at high velocity, but without very significant changes.

To determine the spectrum, we follow the analysis of
\cite{Chang:2008xa}. The most proper analysis would be to include the
updated measurement of $L_{\rm eff}$ (which will weaken the limits) and
the reanalysis to remove multi-hit events from
\cite{Collaboration:2009xb} (which strengthens the limits). However,
incorporating the S2 threshold, resulting in contribution to an
energy-dependent acceptance at low energies is complicated. Lacking
detailed information on this, we will restrict ourselves to the
previous XENON10 limits. Since our point is the relative change from
halo model to halo model, this is reasonable, so long as we recognize
that the limits should be taken with a grain of salt.

We show in Fig.~\ref{fig:paramlight} the allowed parameter space for
the LDM scenario for the canonical MB model, for the (averaged)
simulations VL2, GHALO and GHALO-scaled, as well as two sample spheres from the halos. We see
that the small spectral effects translate into small effects in the
allowed parameter space. There can be some improvement, however, and
although the exclusion curves technically cover the allowed regions,
uncertainties in the low-energy scintillation of Xe, for instance, and
the ambiguity of employing statistical errors without a covariance
matrix suggest that a small region may yet be allowed near 10
GeV.  We should again emphasize
that there is no sense in which the sampling we have done is
statistically complete, and our failure to find an allowed region does
not preclude the possibility that one exists, or might arise in
another simulation. A proper reanalysis using the data from
\cite{Collaboration:2009xb} is warranted.

\section{Discussion}
\label{sec:discussion}

Direct detection of dark matter intimately ties up questions of
astrophysics and particle physics. In recent years, the range of
particle physics models has exploded, many with new interactions and
properties. Some of these, in particular light dark matter and
inelastic dark matter, are particularly sensitive to the high velocity
tail of the WIMP velocity distribution. The same is true, even for
standard WIMPs, for many directionally sensitive detection experiments
(e.g. DRIFT, NEWAGE, DMTPC), which rely on high recoil energies in
order to measure the direction of the scattering DM particle.  It is
especially at these high velocities that deviations from the standard
isothermal Maxwell-Boltzmann assumption are expected to be most
significant. Velocity substructure, arising from nearby subhalos and
tidal streams, as well as from the host halo's velocity anisotropy,
can have profound effects on the expected event rates, the recoil
energy spectrum, and the typical direction of scattering DM particles.

We have studied the effects of these deviations by employing data from
N-body simulations directly, rather than with analytic fits that miss
interesting structures. Our analysis is based on two of the highest
resolution numerical simulations of Galactic dark matter structure,
Via Lactea II and GHALO. We confirm previously reported global
departures from the Maxwell-Boltzmann distribution, with a noticeable
excess of particles on the high velocity tail. In many of our local
samples we find additional discrete features due to subhalos and tidal
streams.

These global and local departures from the standard Maxwellian model
have a number of interesting consequences:

\begin{itemize}
\item For directionally sensitive detection, we have found that the
  fraction of particles coming from the ``hottest'' direction in the
  sky increases with the velocity threshold. More importantly, the
  direction from which most DM particles are coming can be affected
  significantly. At high velocity thresholds ($\gtrsim 500$ km/s in
  the Earth's frame) the typical direction is more often than not
  shifted by $>10^\circ$ away from the direction of Earth's motion,
  and up to $80^\circ$ in extreme cases.

\item For inelastic DM the effects are dramatic. The relative strength
  of CDMS limits versus DAMA can change by an order of magnitude,
  while the relative strength of CRESST limits can change by almost a
  factor of two. For CRESST most of this effect is due to an enhanced
  modulation, which can be a factor of two larger than for a
  Maxwellian halo. Such effects would persist at Xenon detectors as
  well, changing their sensitivity relative to DAMA by
  $\mathcal{O}(1)$.

\item For light DM, we have seen that small changes in the spectrum
  are possible, although not so much as to allow very light ($\sim$ 1
  GeV) WIMPs to fit the existing data, but only somewhat heavier
  ($\gsim$ 10 GeV).

\item In general, dramatic variations in the spectrum can appear for iDM models,
  complicating attempts to employ Yellin-type analyses, which rely
  upon a detailed knowledge of the predicted spectrum of the
  model. Instead we advocate a binned Poisson analysis, which is less
  sensitive to assumption about the shape of the signal spectrum.

\item A further important effect that we have found is the possible
  variation of the phase of modulation as a function of energy. This
  has been noted before in the context of streams
  \cite{Bernabei:2005hj}. Since most background modulations would be
  expected to have the same phase as a function of energy, such a
  change could be a strong piece of evidence that a modulated signal
  was arising from DM.
\end{itemize}

It is important to keep in mind that our simulations are nowhere close
to resolving the phase-space structure at the relevant sub-parsec
scales, and we rely on extrapolation from a coarser sampling (1-1.5
kpc radius spheres). If velocity substructure does not persist on
smaller scales, then our analysis may overestimate the likelihood of
these effects. On the other hand, we are probably underestimating the
significance of the effects, since at the moment we are diluting the
substructure signal by the host halo particles, while it would
completely dominate the background host halo, should the Earth be
passing through one of these substructures.

We have seen here how, should DM be conclusively discovered with
direct detection experiments, it may begin to help unlock questions
about the formation of the galaxy, precisely \textit{because} of these
dramatic sensitivities. With the fantastic improvements in direct
detection on the horizon it is more important than ever to increase
our awareness of the significant impact that the detailed structure of
our Galaxy's dark matter halo may have.

\section*{Acknowledgments}
The authors thank S.Chang and A.Pierce for discussions and their
collaboration in the development of the DM direct detection analysis
tools. MK gratefully acknowledges support from William L. Loughlin at
the Institute for Advanced Study, and from the Theoretical
Astrophysics Center at UC Berkeley. NW is supported by DOE OJI grant
\# DE-FG02-06ER41417 and NSF CAREER grant PHY-0449818.

\bibliography{nbodyidm}
\bibliographystyle{JHEP}
\end{document}